\documentclass[fleqn,usenatbib]{mnras}

\usepackage{newtxtext,newtxmath}
\usepackage[T1]{fontenc}

\DeclareRobustCommand{\VAN}[3]{#2}
\let\VANthebibliography\thebibliography
\def\thebibliography{\DeclareRobustCommand{\VAN}[3]{##3}\VANthebibliography}

\usepackage{graphicx}	% Including figure files
\usepackage{amsmath}	% Advanced maths commands

\usepackage{verbatim}
\usepackage{subcaption}
\usepackage{booktabs}  % For better table formatting

\usepackage{threeparttable}

\title[Stellar multiplicity of exoplanets]{A high stellar multiplicity rate amongst \textbf{\textit{TESS}} planet candidates in the Neptunian desert using \textbf{\textit{Gaia}} DR3 astrometry}

\author[Fintan Eeles-Nolle et al.]{
Fintan Eeles-Nolle$^{1,2}$\thanks{E-mail: \href{mailto:Fintan.Eeles-Nolle@warwick.ac.uk}{Fintan.Eeles-Nolle@warwick.ac.uk}}
and David J. Armstrong$^{1,2}$
\\
$^{1}$Centre for Exoplanets and Habitability, University of Warwick, Coventry, CV4 7AL, UK\\
$^{2}$Department of Physics, University of Warwick, Coventry, CV4 7AL, UK\\
}

\date{Accepted XXX. Received YYY; in original form ZZZ}

\pubyear{2025}

\begin{document}
\label{firstpage}
\pagerange{\pageref{firstpage}--\pageref{lastpage}}
\maketitle

\begin{comment}
    is another of these features, shared by the Neptunian Desert planets and Hot Jupiters.
    An anomalous population of planets has recently been discovered in the previously barren Neptunian Desert. To understand these unusual planets it is important to recognise system and planetary properties that the Desert shares with more populous and well-studied types of exoplanets. 
\end{comment}

\begin{abstract}
We aim to discover whether the stellar multiplicity rate may provide information on the origin of recently discovered planets in the Neptunian Desert. Using {\it Gaia} DR3 astrometry, we search for common proper motion companions to 1779 known exoplanet hosts and 2927 exoplanet candidate hosts from the {\it TESS} mission, both within 650 pc. We find overall stellar multiplicity rates of $16.6\pm0.9\%$ and $19.8\pm0.6\%$ for confirmed and candidate exoplanets, respectively. We find stellar multiplicity rates of $16.7\pm5.8\%$ and $27.5\pm2.6\%$ for confirmed and candidate exoplanets in the Neptunian Desert, respectively. Hot Jupiter host stars were found to have rates of $25.8\pm2.1\%$ and $22.9\pm1.3\%$. For the sample of candidate exoplanets, we find higher stellar multiplicity rates for stars hosting both Hot Jupiters and Neptunian Desert planets compared to control samples of similar stars not known to host planets. For the sample of confirmed exoplanets an increased multiplicity rate is seen for Hot Jupiter hosts, but cannot be significantly determined for Neptunian Desert planet hosts due to small sample sizes. If the candidates from {\it TESS} are indeed planets, the increased multiplicity rate observed could indicate that the Neptunian Desert and Hot Jupiter populations share similar formation mechanisms and environmental conditions. Alternatively, the {\it TESS} candidate high multiplicity rate could imply a prevalence of false positives related to binary and triple stars in this parameter space.
\end{abstract}


%%%%%%%%%%%%%%%%%%%%%%%%%%%%%%%%%%%%%%%%%%%%%%%%%%

%%%%%%%%%%%%%%%%% BODY OF PAPER %%%%%%%%%%%%%%%%%%

\section{Introduction}
\label{sec:intro}

Over the last decades, the number of confirmed exoplanets has exponentially increased, with over 5,800 confirmed planets today on the NASA Exoplanet Archvie\footnote{\url{https://exoplanetarchive.ipac.caltech.edu/}} \citep{exoarch}. With the aid of large exoplanet detection surveys such as Kepler \citep{kepler2010} and the Transiting Exoplanet Survey Satellite ({\it TESS}) \citep{tess2014}, planets have been discovered with a wide range of orbital periods and planetary radii. Additionally, further characterisation work, such as radial velocity or transit spectroscopy observations, has revealed the existence of many different types of exoplanet. As a result, one of the main goals of exoplanetary science has been to understand the different formation mechanisms and environments that can explain the large observed distribution of exoplanet properties - a range much greater than that observed in our own Solar System.

An interesting trend in the distribution of known exoplanets is the so-called hot Neptunian Desert \citep{szabo2011, benitez2011, youdin2011, beauge2013, helled2015, lundkvist2016, mazeh2016, castro2024}: a region consisting of Neptune-sized planets with short orbital periods. This phenomena cannot simply be a result of detection bias, as current methods, such as the transit method, favour larger planets with shorter orbital periods. Significant populations have been discovered at lower radii and longer periods despite this bias. Currently, there exist two possible explanations for this dearth: atmospheric escape and orbital decay. If hot Neptune-sized planets form in-situ, the intense irradiation from their host stars is likely to cause significant atmospheric evaporation, reducing the planetary radii \citep{ehrenreich2011, lopez2014, owen2018}. Additionally, the core accretion model of planet formation suggests that giant planets may form at much larger separations than those associated with the Neptunian Desert, implying planets with these parameters could have formed much further away from their host star and then migrated inwards \citep{pollack1996, lee2015, lee2019}. This orbital decay may occur through disk-driven migration shortly after planet formation \citep{goldreich1979, lin1996, baruteau2016}, or via high eccentricity tidal migration (HEM) at some later period in the planet's lifetime \citep{ford2008, chatterjee2008, beauge2012}.

A key cause of HEM are Kozai-Lidov (K-L) oscillations \cite{wu2003} if the smaller body has a highly inclined orbit. In this case, the transfer of angular momentum between the small inner body and the larger outer object causes the inclination and eccentricity to oscillate synchronously. In the case where this small inner body is another star, the oscillations are a likely cause of the abundance of compact binary stars. Here, K-L oscillations induced by a wide companion star drive the tidal circularisation of an initially long-period binary system towards a more compact form with periods of < 10 days \citep{fabrycky2007}. As a result of this orbital migration, it is likely any additional objects closer to the primary star would be ejected from the system due to instability. In the case where the small inner body is an exoplanet, K-L oscillations may similarly be responsible for the existence of short-period gas giants such as Hot Jupiters and Hot Neptunes.

By its nature, the sample of confirmed exoplanets in the Neptunian Desert is small, and thus to study the nature of the Desert we look to samples of exoplanets that share similar properties, such as a short orbital period. Hot Jupiters, as gas giants and a large population, make for an interesting comparison. Additionally, both Hot Jupiters and Hot Neptunes are known to favour host stars with higher than average metallicities \citep{petigura2018, dai2021}. This similarity in host star properties may hint at further shared features, such as formation or environmental conditions. If Hot Jupiter host system trends are followed by the Hot Neptune population, the much larger population of Hot Jupiters could be used as a statistical probe of the features of Neptunian Desert planets.

One such feature of Hot Jupiters is that they appear to be more common in systems with a bound stellar companion (i.e. the planet is in an S-type orbit) than other types of exoplanet \citep{ngo2016}. This is not the case for all short-period planets, as Super-Earths experience a dearth in multi-star systems \citep{fontanive2021}. If exoplanets in the Neptunian Desert are also found to be more common in multi-star systems, the HEM processes that are predicted to favour Hot Jupiter formation may also contribute to the Desert population in a significant way.

In this work, we investigate the stellar multiplicity of stars hosting confirmed exoplanets and candidates within a 650 pc volume limit. In Section \ref{sec:search samples}, we describe how our samples of (candidate) exoplanet host stars were built. Next, in Section \ref{sec:comp search} we detail the stellar companion search which makes use of the precise astrometry from {\it Gaia} DR3 to find objects with consistent parallaxes and proper motions as our search sample stars. Section \ref{sec:control} outlines how we build control samples of stars with similar spectral distributions as our search samples, which we used to compare the stellar multiplicity of stars hosting (candidate) exoplanets with stars irrespective of their planet-hosting nature. Our results are presented in Section \ref{sec:results}, which we discuss and summarise in Sections \ref{sec:discussion} and \ref{sec:conclusion}.

\section{DR3 Search Samples} \label{sec:search samples}

\subsection{Exoplanet Hosts}\label{sec:search sample}

To begin, we consult the NASA Exoplanet Archive \cite{exoarch} to find a list of confirmed exoplanet systems. This catalogue includes useful data on the exoplanet hosts, such as the corresponding {\it Gaia} DR2 IDs, that make later cross referencing much easier for the majority of the sample. As we are interested in planetary systems with good quality astrometric data in {\it Gaia} DR3, we only include planets that were discovered by primary transits, radial velocities (RVs), eclipse timings, transit timing variations (TTVs), astrometry or direct imaging. The omission of planets detected using microlensing simplifies the {\it Gaia} search, as many of these planets are at large distances or in crowded regions, which would complicate the cone searches. This sample of exoplanets is cut further by removing any planets with: a controversial flag; no radius measurement (including those with only radius limits); no orbital period measurement; and no follow-up transit observations (assuming the initial discovery method was not via transits). Furthermore, we are interested in exoplanets with an outer stellar companion, so planets in circumbinary orbits were not included. This leaves a sample of 4174 exoplanets.

The {\it Gaia} DR3 IDs of the host stars in this sample are found using the \texttt{dr2\_neighbourhood} catalogue within {\it Gaia} DR3, which cross matches {\it Gaia} DR2 IDs with objects in DR3. This is required, especially in the search for companions, as neighbouring stars that were unresolved in DR2 may be resolved in DR3. The ID of an object in one data release does not always correspond to the same ID in other data releases. For hosts with DR2 IDs that matched with more than one DR3 ID, the ID chosen was that of the object with a positive parallax measurement ($\pi > 0$), similar G magnitude ($M_{G, \mathrm{DR3}} - M_{G, \mathrm{DR2}} < 0.1$) and a DR3 sky coordinate that is consistent with the values in DR2, taking into account the proper motion. Since the difference in observational epochs between DR2 and DR3 is only half a year, this last condition is simply 
\begin{equation}
    \mathrm{ang \ sep} < \mu_{\mathrm{DR3}} \ / \ 0.5
\end{equation}
where $\mu_{\mathrm{DR3}}$ is the total proper motion of the DR3 object (the \texttt{dr2\_neighbourhood} catalogue provides pre-calculated angular separations). 50 exoplanet hosts did not have a corresponding DR3 object that followed these parallax, magnitude and coordinate conditions.

35 exoplanets (28 systems) were found to have host stars without {\it Gaia} DR2 IDs in the Exoplanet Archive. Using the alternative hostnames provided in the archive, such as the TIC or HIP IDs, 10 of these systems were found to have {\it Gaia} DR3 IDs and were added to the main sample. The remaining exoplanet systems with no associated {\it Gaia} DR3 IDs were removed.

Now we have the {\it Gaia} DR3 IDs for our sample of exoplanet host stars, we can begin the search for stellar companions. This search involves comparing the parallaxes and proper motions of neighbouring objects, thus we need the search sample to consist of stars that have these quantities to high significance. The parallax and proper motion significances are defined as

\begin{equation}
     \frac{\pi}{\sigma(\pi)} > 3 \quad \& \quad \frac{\mu}{\sigma(\mu)} > 3
    \label{eq:host target uncertainties}
\end{equation}
where $\sigma$ is the error on the respective quantity.

\begin{figure*}
    \centering
    \includegraphics[width=0.32\linewidth]{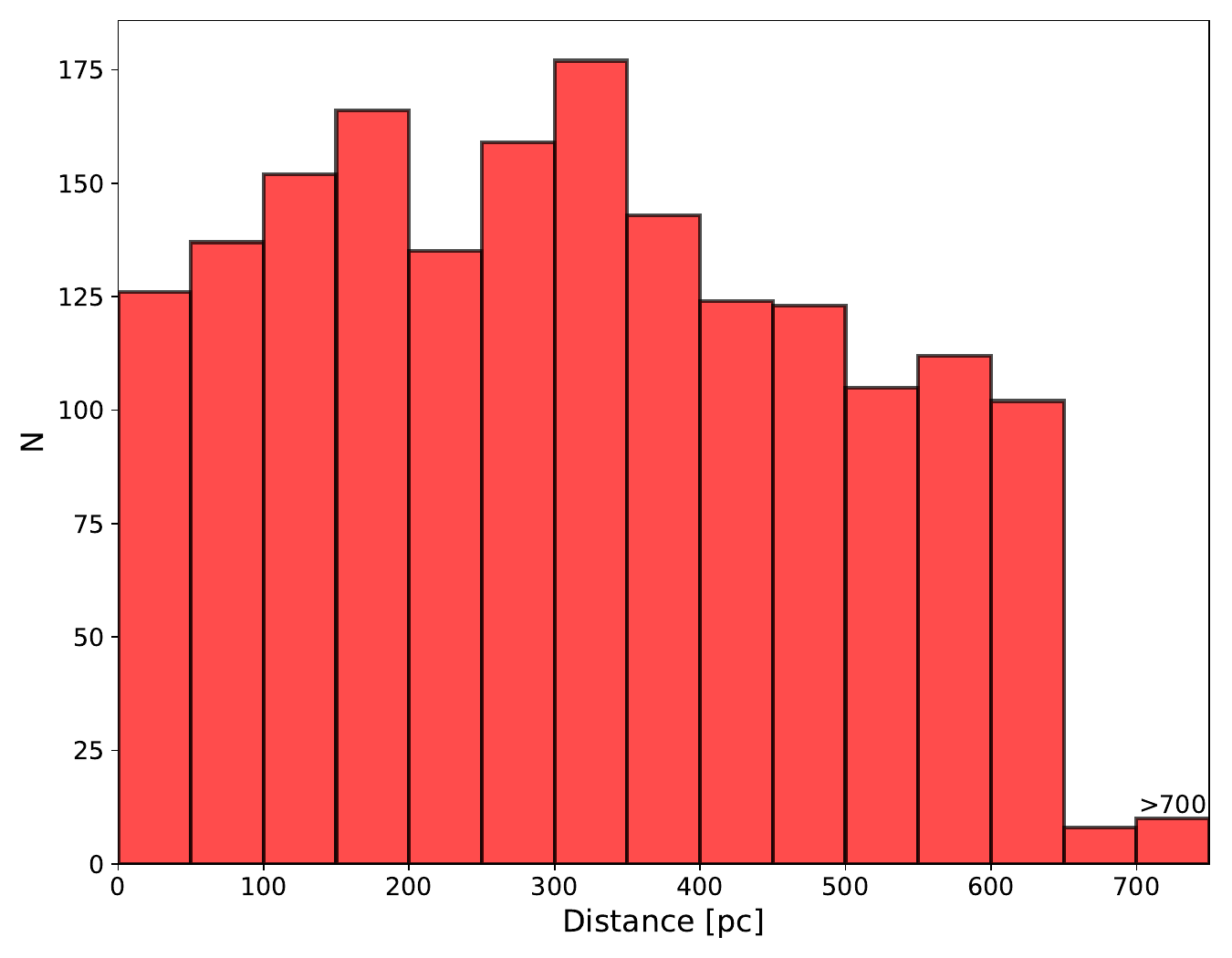}
    \includegraphics[width=0.32\linewidth]{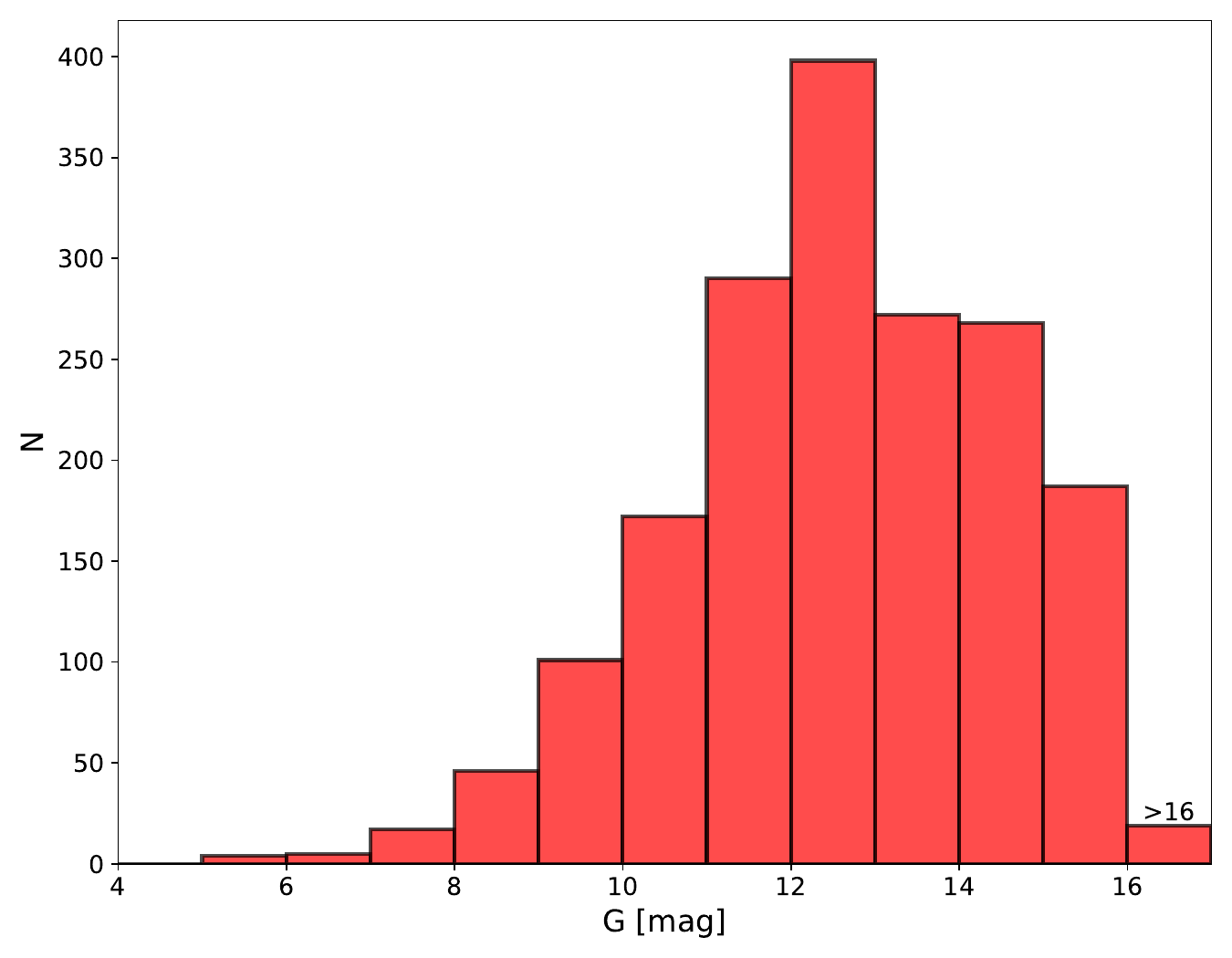}
    \includegraphics[width=0.32\linewidth]{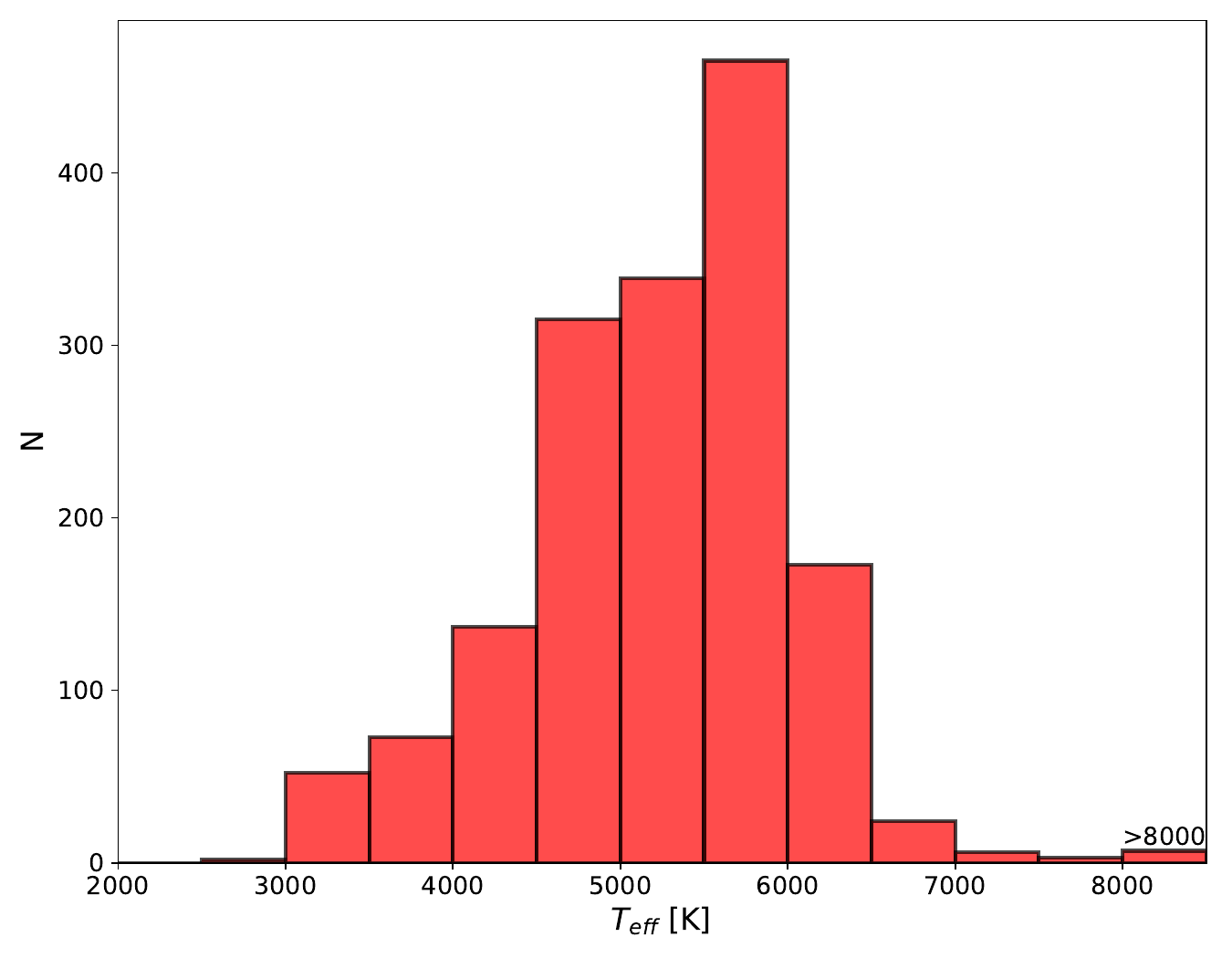}
    \caption{Histograms of the distance (left), G magnitude (centre) and stellar effective temperature (right) of the host stars of confirmed exoplanets included in our search sample. $T_{\mathrm{eff}}$ is derived from the Aeneas algorithm of the Generalized Stellar Parameterizer from Photometry (GSP-Phot) \citep{creevey2023}.}
    \label{fig:search sample}
\end{figure*}

For {\it Gaia} DR3 objects with 5-parameter astrometric solutions, the median parallax uncertainties are 0.02-0.03 mas for $G<15$, 0.07 mas at $G=17$, 0.5 mas at $G=20$, and 1.3 mas at $G=21$ mag \cite{lindegren2021}. Ensuring Equation \ref{eq:host target uncertainties} to hold for even the faintest objects in the search, a parallax limit of $\pi + 3\sigma(\pi) \geq 1.6$ mas is placed on our sample, which corresponds to a volume limit radius of $\sim$ 650 pc around the Sun. Combining the significance conditions and the volume limit gives a search sample of 1779 stars hosting 2372 exoplanets. These targets are listed in Table \ref{tab:search sample}, and histograms showing the distribution of several properties of these targets in {\it Gaia} DR3 are shown in Fig. \ref{fig:search sample}. 

\begin{table*}
    \scriptsize
    \centering
    \begin{threeparttable}
    \caption{Properties of the confirmed exoplanets and their host stars included in our search sample, sorted by DR3 ID. The full table is available online.}
    \label{tab:search sample}
    \begin{tabular}{llrrrlrr}
    \toprule
    Host Name & Gaia ID & Parallax [mas] & Proper Motion [mas/yr] & G [mag] & Planet Name & Orbital Period [days] & Radius [$R_{\oplus}$] \\
    \midrule
    HD 77946 & Gaia DR3 1011435012611767552 & $10.0693\pm0.0235$ & $79.713\pm0.032$ & 8.8599 & HD 77946 b & 6.5273 & $2.705^{+0.086}_{-0.081}$ \\
    HAT-P-13 & Gaia DR3 1014520826353577088 & $4.0750\pm0.0186$ & $35.585\pm0.023$ & 10.4189 & HAT-P-13 b & 2.9163 & $14.258^{+0.729}_{-0.729}$ \\
    HD 80606 & Gaia DR3 1019003226022657920 & $15.1439\pm0.0170$ & $56.967\pm0.021$ & 8.829 & HD 80606 b & 111.4367 & $11.994^{+0.336}_{-0.336}$ \\
    TOI-3785 & Gaia DR3 1044013542142711296 & $12.5465\pm0.0139$ & $46.086\pm0.017$ & 13.634 & TOI-3785 b & 4.6747 & $5.140^{+0.160}_{-0.160}$ \\
    Qatar-9 & Gaia DR3 1048109222955337600 & $4.6633\pm0.0124$ & $31.801\pm0.017$ & 13.673 & Qatar-9 b & 1.5408 & $11.31^{+0.157}_{-0.157}$ \\
    
    \multicolumn{1}{|c|}{\vdots} & \multicolumn{1}{|c|}{\vdots} & \multicolumn{1}{|c|}{\vdots} & \multicolumn{1}{|c|}{\vdots} & \multicolumn{1}{|c|}{\vdots} & \multicolumn{1}{|c|}{\vdots} & \multicolumn{1}{|c|}{\vdots} & \multicolumn{1}{|c|}{\vdots} \\
    \bottomrule
    \end{tabular}
    \end{threeparttable}
\end{table*}

The search sample consists of exoplanet host stars with apparent G-band magnitudes between 5.2 and 18.7, with a median of 12.6 mag. The distribution of distances to the targets is more uniform, with >100 stars included per 50 pc volume slice up to the limit of 650 pc. 18 exoplanet hosts at distances greater than 650 pc are included in this sample, as their corresponding parallax errors are simultaneously large enough to pass the parallax limit and small enough to pass the significance limit in Equation \ref{eq:host target uncertainties}.

The 2372 confirmed exoplanets in our search sample have orbital periods between 0.2 days and $\sim$465 years, with a median of 6.8 days. This orbital period distribution excludes 6 planets in our sample that were discovered by Direct Imaging (HD 206893c,b, HR 8799d,c,b \& COCONUTS-2b) due to their extreme orbital periods when compared with the majority of our sample. The planetary radii very between 0.31$R_{\oplus}$ and 23.5$R_{\oplus}$, with a median of 2.3$R_{\oplus}$.

\subsection{\textbf{\textit{TESS}} Planet Candidates} \label{sec:tess planet candidates}

Despite the discovery of new exoplanets in the deep Neptunian Desert \citep{armstrong2020, jenkins2020, hacker2024}, the population in this region of parameter space remains very small, especially relative to other regions of parameter space (such as Hot Jupiters). Consequently, any comparison involving stars hosting exoplanets in the Neptunian desert or the planets themselves will have very large associated errors on the result. In order to investigate the statistical properties of such systems, we look to exoplanet candidates, specifically those discovered by the {\it TESS} mission. 

These exoplanet candidates (and thus their host systems) are chosen for several reasons. Firstly, the majority of the current population of confirmed exoplanets inside the hot Neptunian desert are those identified, or confirmed, by the \textit{TESS} mission, such as TOI-332b \citep{osborn2023}, TOI-1853b \citep{naponiello2023} and TOI-3261b \citep{nabbie2024}. The large number of stars observed and short observational period for each sector mean \textit{TESS} is much more likely to detect such planets than other observational facilities. The primary science mission of {\it TESS} to focus on nearby bright stars also couples well with \textit{Gaia}, as it ensures most of the host stars of the exoplanet candidates will have good quality astrometric data. Additionally, the host stars are found at small enough distances such that the companion search may have a higher completeness for low-mass stars. This is not necessarily the case for stars hosting exoplanet candidates from the Kepler mission, which are at much larger distances ($\sim800$ pc on average). Searching for nearby stellar companions to these exoplanet candidate host stars would likely produce a much lower number of companions, as the majority of the candidate objects in {\it Gaia} DR3 would not have high enough quality astrometric or photometric data to be classified as a valid companion.

\begin{figure*}
    \centering
    \includegraphics[width=0.32\linewidth]{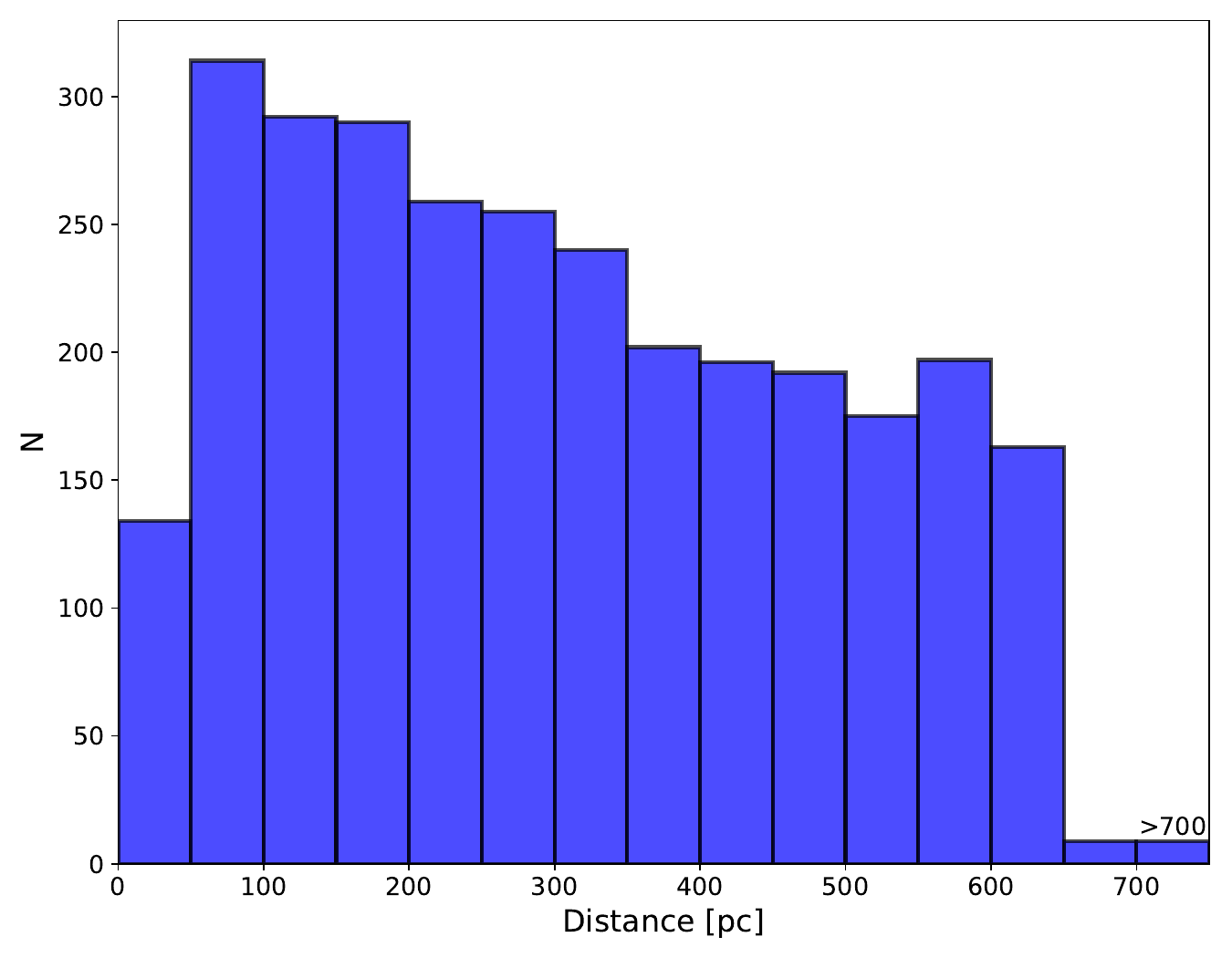}
    \includegraphics[width=0.32\linewidth]{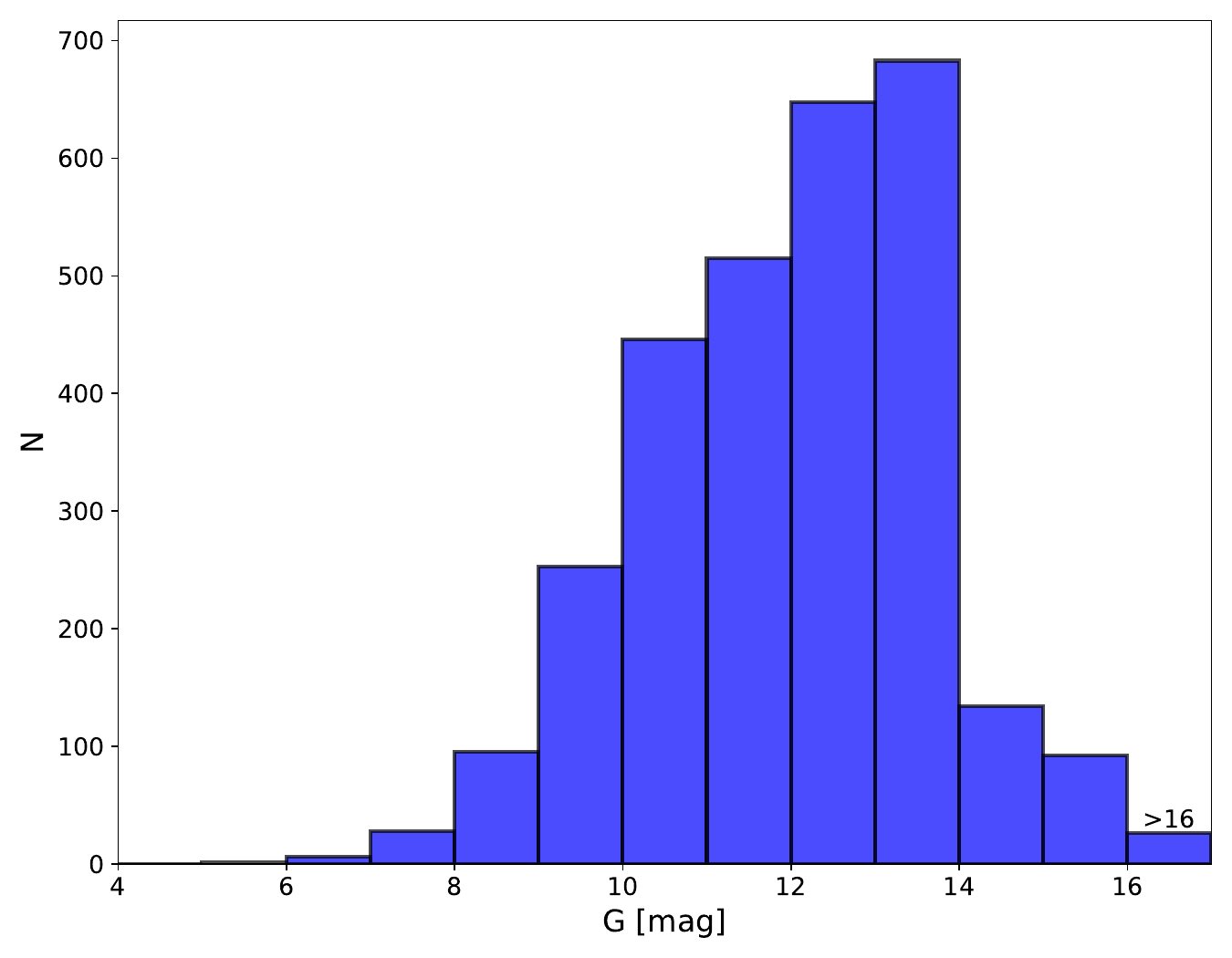}
    \includegraphics[width=0.32\linewidth]{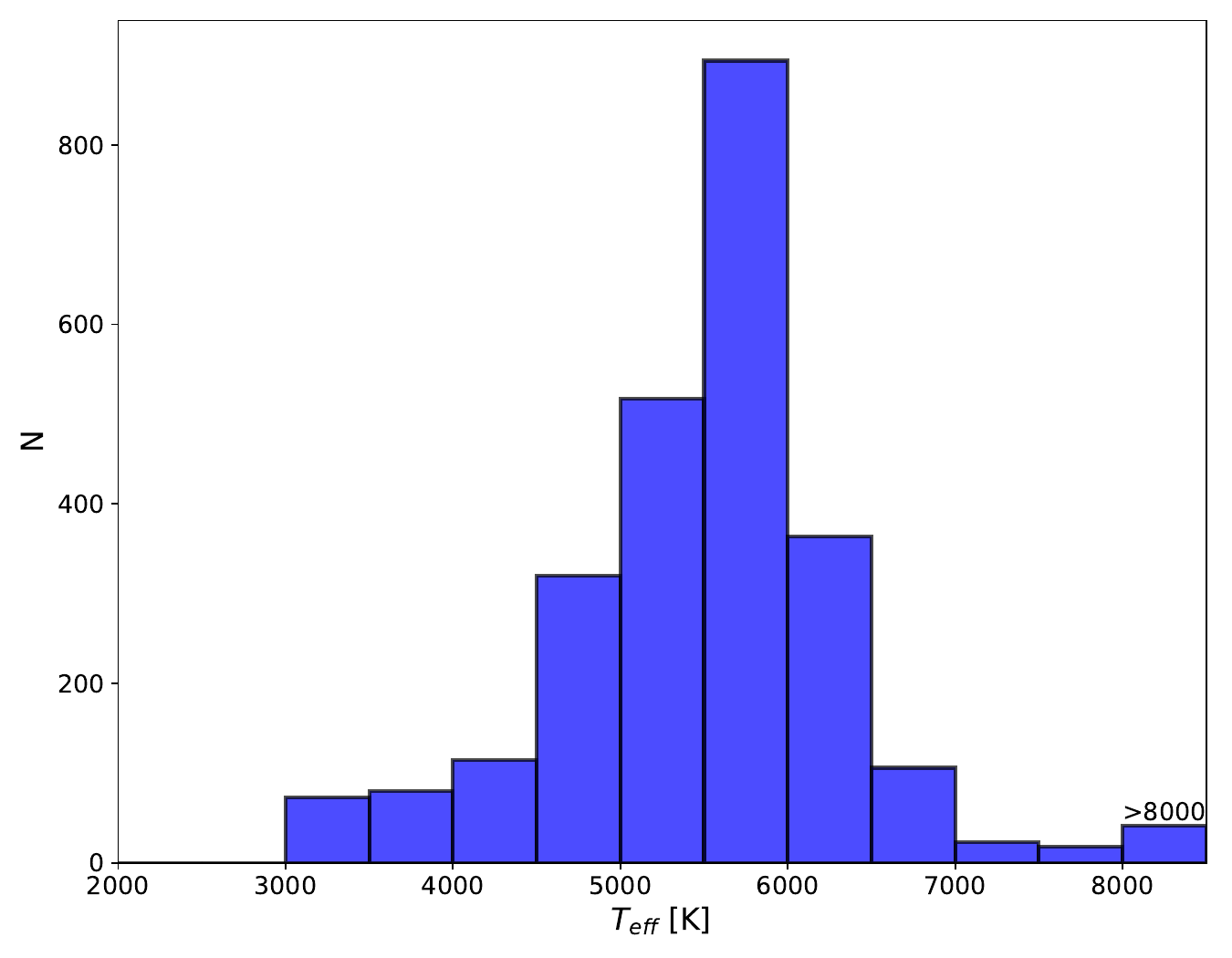}
    \caption{Histograms of the distance (left), G magnitude (centre) and stellar effective temperature (right) of the host stars with associated TOI PCs included in our search sample. $T_{\mathrm{eff}}$ is derived from the Aeneas algorithm of the Generalized Stellar Parameterizer from Photometry (GSP-Phot) \citep{creevey2023}.}
    \label{fig:search sample pc}
\end{figure*}

Using the list of {\it TESS} Objects of Interest (TOIs) provided by the Exoplanet Follow-up Observing Program for {\it TESS} (ExoFOP-TESS)\footnote{\url{https://exofop.ipac.caltech.edu/tess/view_toi.php}}, we retrieve a list of the TOIs that have been given a Planet Candidate (PC) disposition by the {\it TESS} Follow-up Working Group (TFOPWG) \citep{guerrero2021}. We choose to only include this disposition to restrict our sample to the most reliable {\it TESS} candidates, notably excluding ambiguous planet candidates (APCs). This gives a starting sample of 4657 TOIs and 4504 corresponding stars. Using the Mikulski Archive for Space Telescopes (MAST)\footnote{\url{https://mast.stsci.edu/portal/Mashup/Clients/Mast/Portal.html}}, we use the TICv8 catalogue \citep{stassun2019} to find the {\it Gaia} DR2 IDs of our sample. We found 16 stars that had no associated {\it Gaia} IDs in the TIC, which were removed. From this point, we perform the same search process as described in Section \ref{sec:search sample} to find the {\it Gaia} DR3 data for our search sample of {\it TESS} PCs. After applying the same significance conditions and volume limit, we are left with 2927 stars with 3075 associated TOI PCs. This search sample of {\it TESS} planet candidates is listed in Table \ref{tab:search sample pc}. Fig. \ref{fig:search sample pc} shows the distribution of {\it Gaia} DR3 properties in this sample.

\begin{table*}
    \scriptsize
    \centering
    \begin{threeparttable}
    \caption{Properties of the TOIs and the associated host stars included in our search sample of exoplanet candidates from {\it TESS}, sorted by DR3 ID. The full table is available online.}
    \label{tab:search sample pc}
    \begin{tabular}{llrrrcrr}
    \toprule
    TIC ID & Gaia ID & Parallax [mas] & Proper Motion [mas/yr] & G [mag] & TOI & Orbital Period [days] & Planet Radius [$R_{\oplus}$] \\
    \midrule
    14336130 & Gaia DR3 1000423609817438592 & $9.6247 \pm 0.0199$ & $75.362 \pm 0.019$ & 9.2485 & 1716.01 & 8.0823 & $2.773 \pm 0.183$ \\
    88385463 & Gaia DR3 1002542261348598528 & $2.8408 \pm 0.0129$ & $46.432 \pm 0.016$ & 12.8523 & 3822.01 & 3.1235 & $13.291 \pm 0.737$ \\
    88565745 & Gaia DR3 1002774052143638784 & $3.2029 \pm 0.0220$ & $4.475 \pm 0.026$ & 11.6043 & 5571.01 & 731.4779 & $11.778 \pm 2.172$ \\
    306401382 & Gaia DR3 100311164516850944 & $4.0972 \pm 0.0263$ & $34.776 \pm 0.039$ & 14.6374 & 5311.01 & 39.0537 & $13.525 \pm 0.823$ \\
    70897115 & Gaia DR3 1006506898417155712 & $1.7034 \pm 0.0157$ & $5.822 \pm 0.013$ & 13.6687 & 6988.01 & 13.4140 & $8.766  $ \\
    \multicolumn{1}{|c|}{\vdots} & \multicolumn{1}{|c|}{\vdots} & \multicolumn{1}{|c|}{\vdots} & \multicolumn{1}{|c|}{\vdots} & \multicolumn{1}{|c|}{\vdots} & \multicolumn{1}{|c|}{\vdots} & \multicolumn{1}{|c|}{\vdots} & \multicolumn{1}{|c|}{\vdots} \\
    \bottomrule
    \end{tabular}
    \end{threeparttable}
\end{table*}

The search sample of {\it TESS} exoplanet candidates contains stars with apparent G-band magnitudes between 5.4 and 17.2, with the highest frequency a median of 12 mag. Apart from the first bin, each 50 pc bin has > 150 targets up to the volume limit, with a median distance of 284 pc. This sample also includes 18 targets at distances greater than the 650 pc volume limit.

The exoplanet candidates in this sample have orbital periods between 0.2 and $\sim$1850 days, with a median of $\sim 5$ days. This excludes 73 TOIs with no listed orbital period. The planetary radii vary between 0.57$R_{\oplus}$ and 45.4$R_{\oplus}$, with a median of 7.6$R_{\oplus}$.

\section{Companion Search}\label{sec:comp search}

With our search samples of stars hosting confirmed exoplanets and {\it TESS} planet candidates, we can now search for nearby stellar companions. To do this, a cone search around each target is performed in the full \texttt{gaia\_source} catalogue of DR3. We expect to find companions across a wide range of separations, with distributions strongly dependent on the mass of the primary star \citep{dhital2010}. As a result of the Gaia spatial resolution of $\sim0.4$ arcsec for close binaries \citep{lindegren2018}, we do not expect to detect companions at separations $\lesssim100$ AU. For companions at much larger separations, \citet{el-badry2021} found that chance alignments begin to dominate the sample of companions beyond 30,000 AU, even with strong data quality cuts. Thus, we employ an approximate outer companion separation limit of 10,000 AU. To search this region around the target object, we use a cone search with radius that depends on the parallax of the target host star, $r = 10 \pi $, where $r$ is in arcsec and $\pi$ is in mas \citep{mugrauer2019}. Of the objects found using this cone search, only those with significant parallax and proper motion measurements are retained (ie. the same conditions as Equation \ref{eq:host target uncertainties} but for the companion candidate objects).

To find the bound companions in the surviving sample, each object is then assessed for common proper motion (CPM) with the target host star. In this case, we use two conditions that represent the likelihood of companionship. Firstly, bound companions should have a common proper motion. We use the common proper motion index (CPM-Index) from \citet{mugrauer2019} and define systems with 
\begin{equation}
    \frac{|\mu_{\mathrm{host}} + \mu_{\mathrm{comp}}|}{|\mu_{\mathrm{host}} - \mu_{\mathrm{comp}}|} > 3 \ 
\end{equation}
as {\it Gaia} DR3 objects that are a common proper motion pair.
Secondly, companion objects should appear to be at similar distances as their corresponding target star. Considering the parallax difference between the two objects, companions are defined to have insignificant differences, ie. 
\begin{equation}
    \frac{\Delta\pi}{\sqrt{\sigma^{2}(\pi_{\mathrm{comp}}) + \sigma^{2}(\pi_{\mathrm{host}})}} < 3 \ ,
    \label{eq:parallax diff sig}
\end{equation}
where $\Delta\pi = |\pi_{\mathrm{comp}} - \pi_{\mathrm{host}}|$.
This assessment is valid for the majority of the sample, as at distances >100 pc any stellar companions should appear to have consistent parallax measurements with the host star. However, this is not necessarily the case for close, bright stars in regions of low stellar density. The astrometric precision of such objects in {\it Gaia} DR3 is high enough that a line of sight separation of $\sim$10,000 AU between two stars could not result in parallax values consistent with each other (to 3$\sigma$). It follows that these systems would be removed under Equation \ref{eq:parallax diff sig}, despite the low separation. We therefore modify Equation \ref{eq:parallax diff sig} so that the parallax difference is consistent with a non-zero parallax projection at the parallax of the target star:
\begin{equation}
    \Delta\pi - 3 \sqrt{\sigma^{2}(\pi_{\mathrm{comp}}) + \sigma^{2}(\pi_{\mathrm{host}})} < \pi_{\mathrm{sep}} \ ,
    \label{eq:parallax diff allowed}
\end{equation}
where $\pi_{\mathrm{sep}} [\mathrm{mas}] = 5\times10^{-5} \times (\pi_{\mathrm{host}}[\mathrm{arcsec}])^{2}$ is the maximum parallax difference from a separation of 10,000 AU along the line of sight direction. This condition is optimistic, as multi-star systems with line of sight separations close to this would likely be blended in the {\it Gaia} data. Additionally, this is dependent on the $\pi^{-1}$ [arcsec] approximation of distance being accurate without significant correction. This should not be a limiting factor however, as at the distances where some correction becomes required, the parallax precision on the average DR3 object would be too low to identify such a small line of sight separation.

\section{Control Sample} \label{sec:control}

In order to assess whether the stellar multiplicity rate of our sample is linked in any way to the exoplanet hosting nature of the stars used, we need some way to compare the stellar multiplicity rate of stars hosting exoplanets to that of stars without exoplanets, while avoiding new biases. To do this, we create a control sample of stars to which we perform the same companion search method as described in Section \ref{sec:comp search}.

Additionally, as discussed in Section \ref{sec:tess planet candidates}, one of the main problems with any analysis involving the Neptune Desert is the low confirmed population. Consequently, we expect large errors on any stellar multiplicity rate calculated using this population. In order to evaluate this result and any differences with the general field star population with reasonable significance, we need a larger control sample to reduce our uncertainties. We therefore make our control samples 5 times larger than our search samples, but with the same distributions of stellar parameters and distances. Some stars in our search samples had combinations of parameters and distances that were difficult to find matches for. As a result, the control sample sizes are slightly less than 5 times the search sample sizes. This factor was chosen as the largest value where >99\% of our sample returned 

\subsection{Stellar Parameters}\label{sec:control stellar params}

The selection of the stars to be used in our control sample cannot be random. The non-uniformity of binarity (and higher order multiplicity) across different stellar populations means we could not simply select stars from {\it Gaia} DR3 at random for comparison. Higher mass stars are more likely to have a stellar companion \citep{moe2017}. Consequently, the first restriction we face is that of the stellar properties of both of the search samples - a control sample star should have similar physical properties to the corresponding exoplanet (candidate)-hosting star in the search samples. Binary (and higher order multiplicity) star distribution is usually described in terms of spectral types, with \citep{raghavan2010} finding that nearly half of all nearby ($d<25$pc) solar-type stars have stellar companions. This companion rate decreases significantly for dwarf stars, with estimates of 11 - 42\% for M-dwarfs (depending on the separation) and 10-30\% for L and brown dwarfs \citep{duchene2013}. 

This would be problematic if we were to sample control stars independently of stellar properties, as both the search samples consist largely of G-type stars (35\% and 39\% of the confirmed and TOI samples respectively). This distribution of spectral types is not reflected in {\it Gaia} DR3, with only $\sim 5\%$ of the \texttt{astrophysical\_parameters} catalogue within 650 pc being G-types. Consequently, if we were to randomly sample control stars from this catalogue or another within {\it Gaia} DR3, we would find a significantly different stellar distribution to our search samples.

To reproduce the stellar population distribution of our search samples in the respective control samples, we identified stars in the \texttt{astrophysical\_parameters} catalogue of DR3 with stellar parameters similar to each search sample star:
\begin{itemize}
    \item $T_{*} - 100 < T_{\mathrm{c}} < T_{*} + 100$
    \item $R_{*} - 0.1 < R_{\mathrm{c}} < R_{*} + 0.1$
    \item $[M/H]_{*} - 0.1 < [M/H]_{\mathrm{c}} < [M/H]_{*} + 0.1$
\end{itemize}
where $T_{*}$, $R_{*}$ and $[Z_{*}]$ are the effective temperature, stellar radii and metallicity of the search sample star respectively from the {\it Gaia} DR3 GSP-Phot module \citep{creevey2023}. 

About $11\%$ of the confirmed exoplanet hosts and $16\%$ of the exoplanet candidate hosts failed this search, either by not possessing parameters from GSP-Phot or by returning less than 5 control stars each. For these stars, we used the G, G$_{\mathrm{BP}}$ and G$_{\mathrm{RP}}$ band photometry from {\it Gaia} DR3 as a metric for finding similar stars. We then identified stars in the \texttt{gaia\_source} catalogue of DR3 that have: 
\begin{itemize}
    \item $M_{G,*} - 0.1 < M_{G,\mathrm{c}} < M_{G,*} + 0.1$
    \item $G_{\mathrm{BP},*} - G_{\mathrm{RP},*} - 0.1 < G_{\mathrm{BP},\mathrm{c}} - G_{\mathrm{RP},\mathrm{c}} < G_{\mathrm{BP},*} - G_{\mathrm{RP},*} + 0.1$
\end{itemize}
where $G_{\mathrm{BP},*}$ and $G_{\mathrm{RP},*}$ are the apparent magnitudes of the search sample star in the blue and red \textit{Gaia} bands respectively, and $M_{G,*}$ is the absolute $G$ magnitude of the search sample star. The control samples found using GSP-Phot parameters and photometry were then merged.

\begin{figure*}
    \centering
    \begin{subfigure}{\textwidth}
    \includegraphics[width=0.98\linewidth]{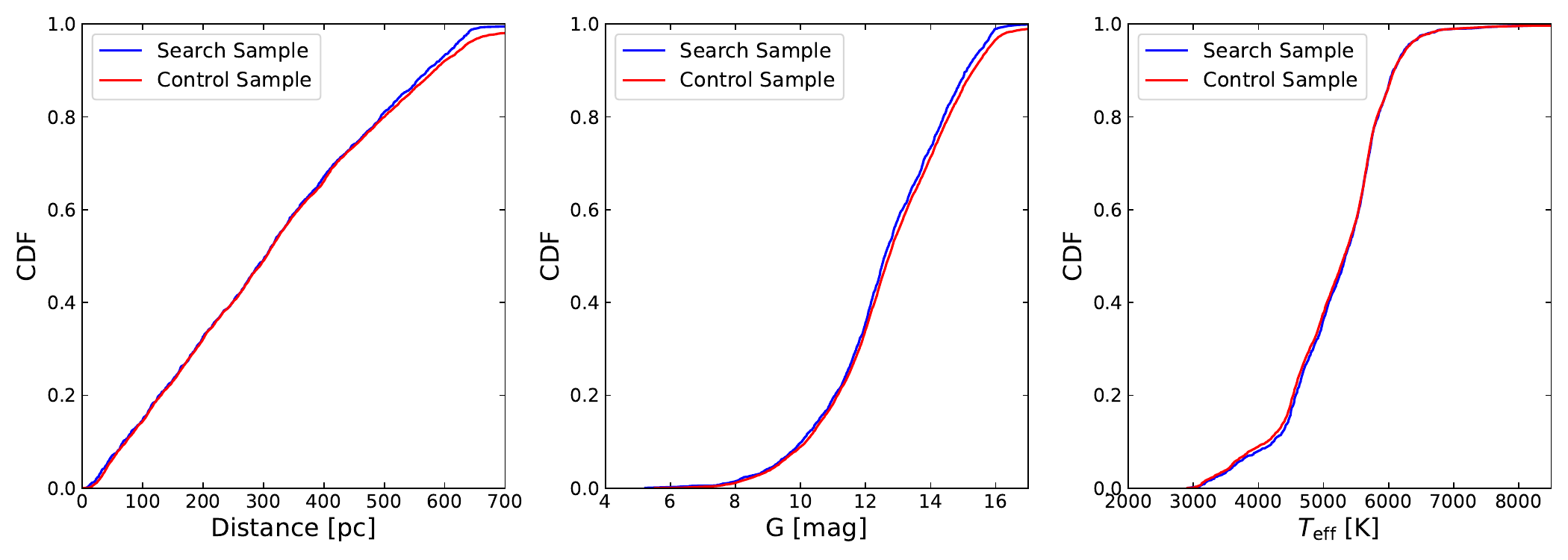}
    \end{subfigure}
    \begin{subfigure}{\textwidth}
    \includegraphics[width=0.98\linewidth]{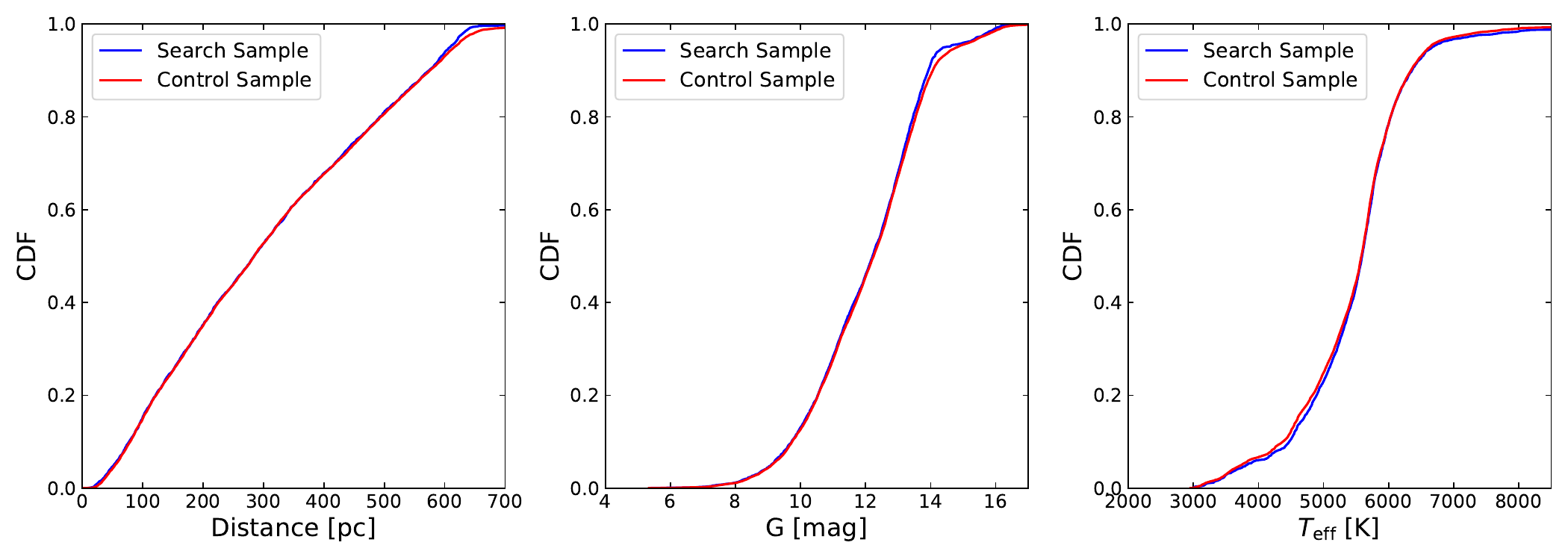}
    \end{subfigure}
    \caption{CDFs of the distance (left), G magnitude (centre) and stellar effective temperature (right) of the search samples of stars hosting confirmed exoplanets (top) and planet candidates (bottom) and their respective control samples. $T_{\mathrm{eff}}$ is derived from the Aeneas algorithm of the Generalized Stellar Parameterizer from Photometry (GSP-Phot) \citep{creevey2023}.}
    \label{fig:control samples}
\end{figure*}

As expected, the control samples have similar distributions of stellar properties as the search samples. This is shown by the cumulative distribution functions in Fig \ref{fig:control samples}, which compare the distance, G mag and $T_{\mathrm{eff}}$ of the search and control samples for both confirmed and candidate exoplanet hosts.

\subsection{RUWE}\label{sec:control ruwe}

The renormalised unit weight error (RUWE) of an object in {\it Gaia} DR3 is a representation of how well it fits a single source astrometric model \citep{gaia2021, lindegren2021b}, such that a perfect fit is close to unity and increasing RUWE values correspond to decreasing fit quality. One of the most probable causes of a large RUWE is that the source is a close binary or a system or higher order multiplicity, as the orbital motions of the blended stars cause the poor fit \citep{belokurov2020, penoyre2020}. Consequently, RUWE is commonly used as a simple indicator of potential unresolved stellar multiplicity in \textit{Gaia} \citep{berger2020, ziegler2020, kervella2022}. Although the distribution of RUWE is larger for stars with larger parallax errors, the peak of the distribution does not experience the same level of parallax dependence. We can therefore simply define stars with RUWE > 1.4 as 'high RUWE' and vice versa, as in \citep{kervella2022}. We then require the control sample star to match the RUWE condition of the corresponding search sample star, ie.
\begin{itemize}
    \item $\mathrm{RUWE}_{\mathrm{c}} < 1.4$ if $\mathrm{RUWE}_{*} < 1.4$
    \item $\mathrm{RUWE}_{\mathrm{c}} > 1.4$ if $\mathrm{RUWE}_{*} > 1.4$
\end{itemize}
where the $*$ and c denote the search and control samples respectively.

Introducing this RUWE condition on the control samples will ensure that stars in the search samples with 'high RUWE' will have corresponding stars in the control samples also with 'high RUWE'. Our goal is to ensure the RUWE distribution remains the same in both samples so we avoid introducing additional unresolved stellar multiplicity into our control sample. This is necessary because a wide companion search around a search sample target with 'high RUWE' may be, in reality, a search for an outer member of a hierarchical triple system. These are much rarer than wide binaries \citep{raghavan2010}, and our search would produce a lower perceived number of detected stellar companions.

\subsection{Distance}\label{sec:control dist}

We place another constraint on the distances of the control sample stars. As the companion search is parallax dependent, building the control sample with a similar distance distribution will ensure that that any biases are present in both the real and control samples. This constraint takes two forms, depending on whether the search sample star has astrophysical parameters derived by the GSP-Phot module. Firstly, for search sample stars that have a distance measurement from the GSP-Phot module, we are interested in stars that are within $\pm 10$pc of the search sample star. For search sample stars that do not have a distance calculated by GSP-Phot, we convert the range of physical distances to a parallax range. In this case, the control sample stars are required to have parallaxes of
\begin{itemize}
    \item $\pi_{*} - 10 \pi_{*}^{2} < \pi_{\mathrm{c}} < \pi_{*} + 10 \pi_{*}^{2}$
\end{itemize}
where $\pi_{*}$ and $\pi_{\mathrm{c}}$ are the parallaxes (in arcsec) of the stars in the search sample and control sample respectively.

With control samples for both our search samples, we then perform the same companion search around each star, as described in Section \ref{sec:comp search}.

\section{Planet Types in Period-Radius Space}\label{sec:planet types}

\begin{figure*}
    \centering
    \includegraphics[width=0.49\linewidth]{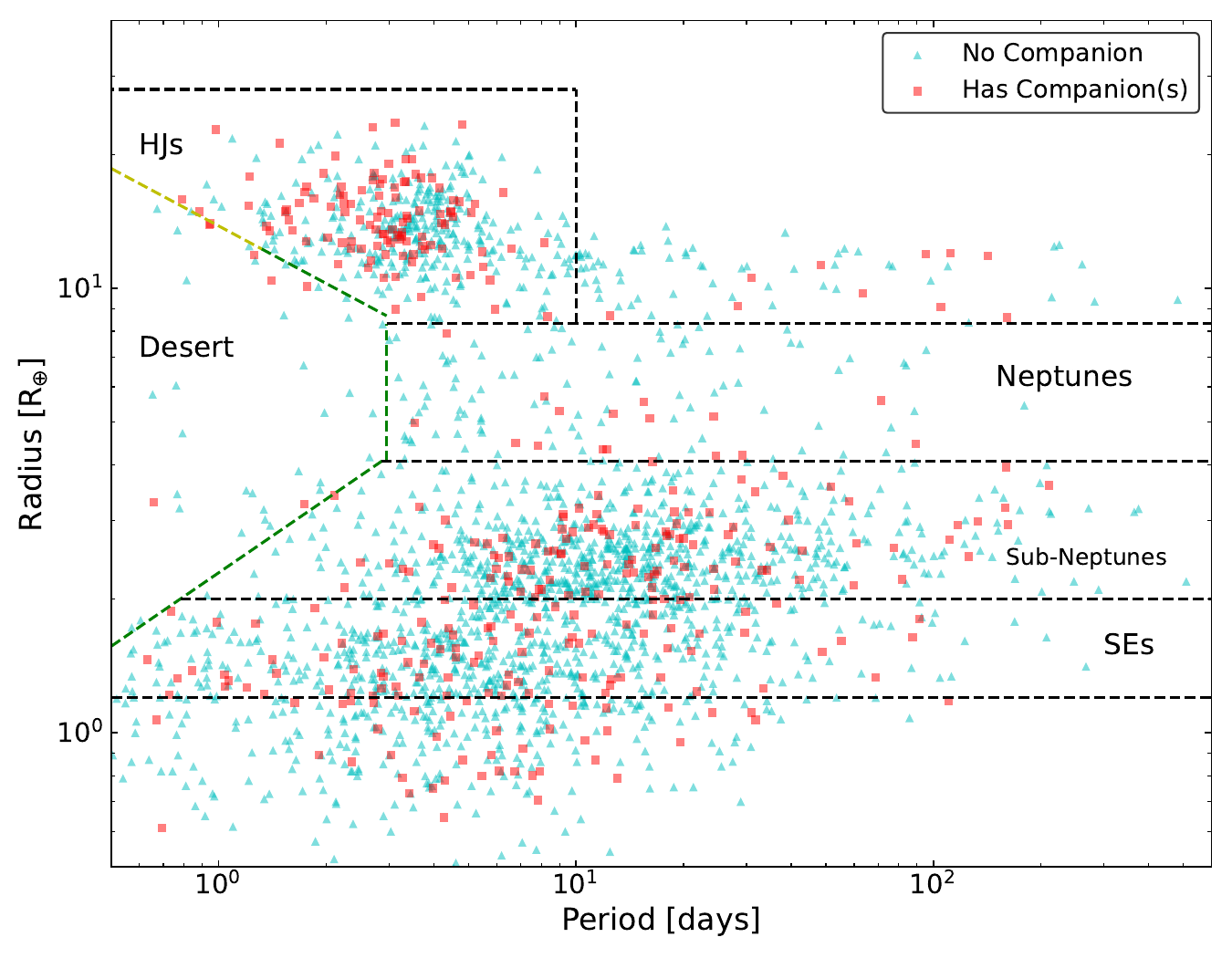}
    \includegraphics[width=0.49\linewidth]{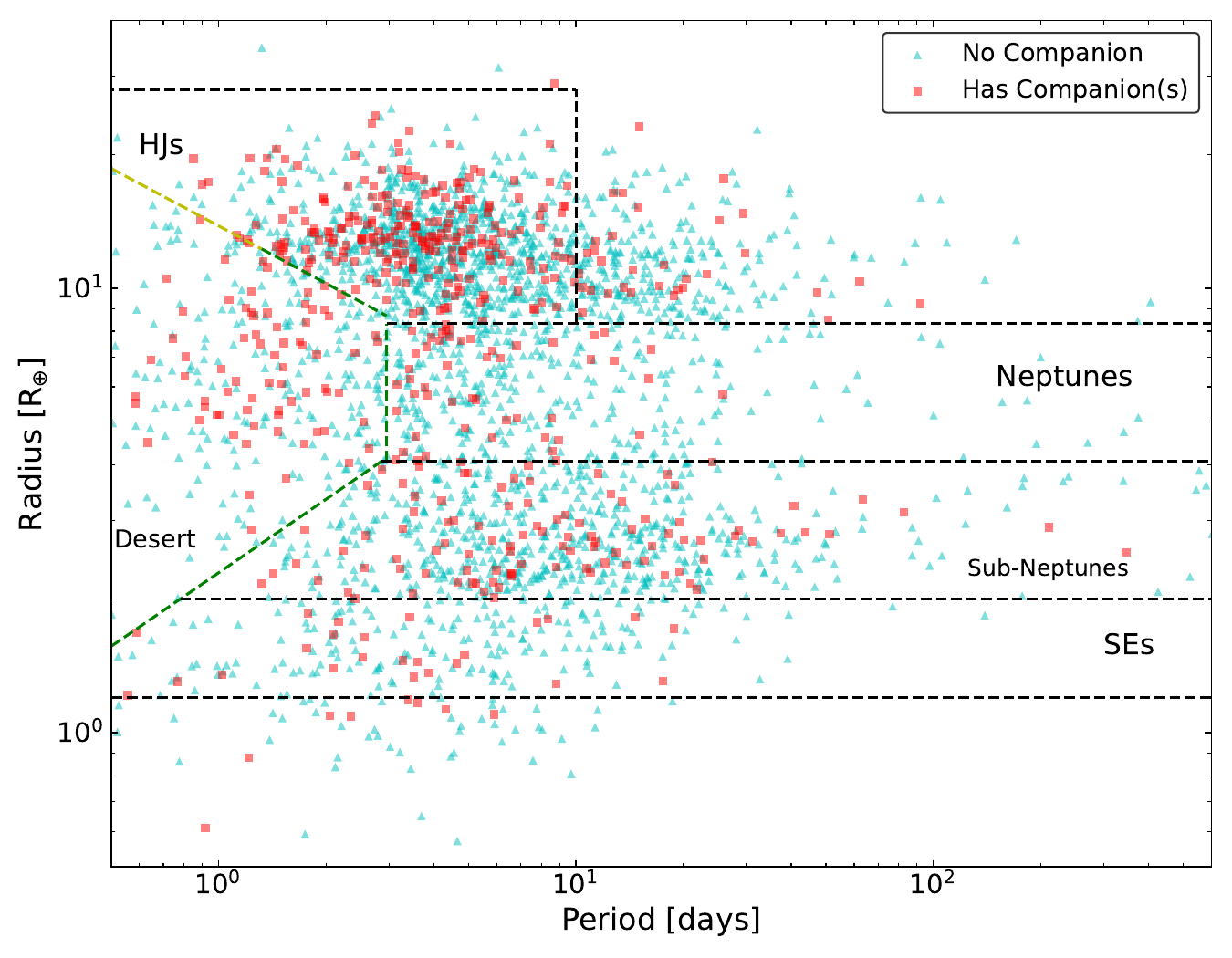}
    \caption{The orbital periods and planetary radii of the current population of confirmed exoplanets (left) and TOIs (right). The regions of period-radius parameter space that are used to define the different types of planet in this work are included.}
    \label{fig:types}
\end{figure*}

In this work, we are principally interested in whether stellar multiplicity appears dependent on the position of exoplanets in period-radius parameter space. In addition to finding the stellar multiplicity rate across the full exoplanet (\& candidate) population, we split the period-radius space into smaller regions, each roughly corresponding to a known 'type' of exoplanet. The main motivation for this is to analyse the planets (\& candidates) in the hot Neptunian desert, which is typically described by empirical boundaries in the period-radius space \citep{mazeh2016, castro2024}. Although the other types of exoplanets we have included are not necessarily defined by specific radii or orbital periods, we label planets in these regions as the respective planet 'type' for comparison purposes only. These definitions should not be assumed to be complete. The parameter space regions are mutually exclusive. A perfect example of this is the Super-Earth region, which, as the Neptune desert lower boundary intersects both the upper and lower radius limits, inhibits any planet in this radius range with an orbital period below $\sim 1$ day from being labeled as a Super-Earth. However, the regions are not collectively exhaustive, as shown in Fig. \ref{fig:types}. This approach allows us to compare the stellar multiplicity in the hot Neptunian desert with that of Hot Jupiters, as described in Section \ref{sec:intro}. 

We define the following regions and the associated planet types:
\begin{enumerate}
    \item Neptunian desert: 
    \begin{itemize}
        \item $0.55\mathcal{L}_{P} + 0.36 < \mathcal{L}_{R} < -0.43\mathcal{L}_{P} + 1.14$
        \item $-0.3 < \mathcal{L}_{P} < 0.47$
    \end{itemize}
    \item Hot Jupiters:
    \begin{itemize}
        \item $0.92 < \mathcal{L}_{R} < 1.45$
        \item $\mathrm{min}(-0.3, -0.43\mathcal{L}_{P} + 1.14) < \mathcal{L}_{P} < 1$
    \end{itemize}
    \item Neptunes:
    \begin{itemize}
        \item $0.61 < \mathcal{L}_{R} < 0.92$
        \item $\mathcal{L}_{P} > 0.47$
    \end{itemize}
    \item Sub-Neptunes:
    \begin{itemize}
        \item $0.3 < \mathcal{L}_{R} < 0.61$
        \item $\mathcal{L}_{P} > 0.55\mathcal{L}_{P} + 0.36$
    \end{itemize}
    \item Super-Earths:
    \begin{itemize}
        \item $0.08 < \mathcal{L}_{R} < 0.3$
        \item $\mathcal{L}_{P} > 0.55\mathcal{L}_{P} + 0.36$
    \end{itemize}
\end{enumerate}
where $\mathcal{L}_{P} = \mathrm{log}_{10}(P_{\mathrm{orb}}/\mathrm{day})$ and $\mathcal{L}_{R} = \mathrm{log}_{10}(R_{\mathrm{p}}/R_{\odot})$. These regions are shown in Fig. \ref{fig:types}.

Firstly, while \citet{mazeh2016} defined the hot Neptunian desert as the region between a pair of upper and lower log-linear boundaries in period-radius space, the lower boundary especially has become less clear with the confirmation of large numbers of sub-Neptunes from the Kepler \citep{kepler2010} and K2 \citep{k22014} surveys. To account for this, we make use of the boundaries presented by \citet{castro2024}, which describes the Neptunian desert using altered upper and lower log-linear boundaries, and with the addition of a third boundary, the "Neptunian ridge", at a constant orbital period at Neptunian radii ranges.

The Hot Jupiters occupy a region of the parameter space above the Neptunian desert. To keep our exclusivity, we choose to use the upper boundary of the Neptunian desert as part of the lower boundary of the Hot Jupiter region in both period and radius. The upper orbital period and radius of this region are defined as $P = 10$ days and $R = 2.5 R_{J}$ respectively.

The longer-period Neptunes are an interesting class of exoplanet to include, as the "Neptunian Ridge" introduced by \citet{castro2024} implies there may exist some physical phenomena separating the short and long period Neptunes. This is not known to occur at other planetary radii. Consequently, we include Neptunes as planets with orbital period $P \gtrsim 4.5 $ days and radii in the range $3R_{\odot} < R < 0.8R_{J} (8R_{\odot})$.

Planets with an orbital period longer than the lower boundary of the Neptunian desert and with radii in the range $1.2 < R/R_{\odot} < 2$ are defined as Super-Earths. We also define sub-Neptunes as planet with radii in the range $2 < R/R_{\odot} < 10^{0.61}$, similarly with orbital periods above the lower Neptunian desert boundary. The upper radius limit corresponds to the lower radius limit of the "Neptunian Ridge" \citep{castro2024}.

\section{Results} \label{sec:results}

In this work, we searched for companion objects to stars hosting confirmed exoplanets or TOIs using {\it Gaia} DR3. From our samples of stars hosting confirmed exoplanets and planet candidates, we find 295 and 579 multi-star systems with 307 and 639 companions respectively. Tables \ref{tab:comps pl} and \ref{tab:comps pc} list the astrometric data of the companion objects and their corresponding exoplanet or candidate host star. This translates to a stellar multiplicity rate of 16.6±0.9\% for confirmed exoplanet hosts and 19.8±0.7\% for planet candidate hosts.

Within these multi-stellar systems, we find 396 confirmed and 658 candidate exoplanets. The orbital periods and radii of the exoplanets or candidates with stellar companions are shown in Fig. \ref{fig:types}.

\begin{table*}
    \scriptsize
    \centering
    \begin{threeparttable}
    \caption{Astrometric properties of the detected companions to exoplanet host stars in {\it Gaia} DR3, sorted by the host star name. The full table is available online.}
    \label{tab:comps pl}
    \begin{tabular}{llrrcr}
    \toprule
    Host & Companion DR3 ID & $\rho$ [arcsec] & $\Delta\pi$ [mas] & Relative Proper Motion [mas/yr] & CPM-Index \\
    \midrule
    55 Cnc & Gaia DR3 704966762213039488 & $84.826207 \pm 0.000045$ & $0.2078 \pm 0.0640$ & $0.850 \pm 0.709$ & 1269.498 \\
    CoRoT-2 & Gaia DR3 4287820852697823872 & $4.080453 \pm 0.000031$ & $0.0622 \pm 0.0425$ & $0.606 \pm 0.036$ & 37.301 \\
    CoRoT-9 & Gaia DR3 4285571659911185280 & $11.968002 \pm 0.000102$ & $-0.0745 \pm 0.1220$ & $0.413 \pm 0.113$ & 101.444 \\
    DS Tuc A & Gaia DR3 6387058411482257280 & $5.365593 \pm 0.000018$ & $0.0156 \pm 0.0195$ & $2.168 \pm 0.021$ & 95.266 \\
    GJ 3473 & Gaia DR3 3094290019967631360 & $49.259780 \pm 0.000048$ & $-0.1623 \pm 0.0695$ & $2.875 \pm 0.094$ & 384.818 \\
    \multicolumn{1}{|c|}{\vdots} & \multicolumn{1}{|c|}{\vdots} & \multicolumn{1}{|c|}{\vdots} & \multicolumn{1}{|c|}{\vdots} & \multicolumn{1}{|c|}{\vdots} & \multicolumn{1}{|c|}{\vdots} \\
    \bottomrule
    \end{tabular}
    \end{threeparttable}
\end{table*}

\begin{table*}
    \scriptsize
    \centering
    \begin{threeparttable}
    \caption{Astrometric properties of the detected companions to stars hosting exoplanet candidates from {\it TESS} in {\it Gaia} DR3, sorted by TIC ID. The full table is available online.}
    \label{tab:comps pc}
    \begin{tabular}{llrrcr}
    \toprule
    TIC ID & Companion DR3 ID & $\rho$ [arcsec] & $\Delta\pi$ [mas] & Relative Proper Motion [mas/yr] & CPM-Index \\
    \midrule
    100757807 & Gaia DR3 2890519660294833792 & $4.331277 \pm 0.000023$ & $0.0041 \pm 0.0252$ & $1.166\pm 0.054$ & 49.952 \\
    100909102 & Gaia DR3 1910906408272899200 & $2.922635 \pm 0.000026$ & $-0.0571 \pm 0.0306$ & $0.111 \pm 0.071$ & 613.314 \\
    101520163 & Gaia DR3 3434583646083607680 & $12.549484 \pm 0.000119$ & $0.0053 \pm 0.1079$ & $0.969 \pm 0.143$ & 30.086 \\
    101696403 & Gaia DR3 6668794109886705408 & $2.493523 \pm 0.000101$ & $-0.1862 \pm 0.1588$ & $0.216 \pm 0.331$ & 270.098 \\
    101948569 & Gaia DR3 4982951791883929600 & $2.348047 \pm 0.000037$ & $-0.0980 \pm 0.0596$ & $0.342 \pm 0.348$ & 172.708 \\
    \multicolumn{1}{|c|}{\vdots} & \multicolumn{1}{|c|}{\vdots} & \multicolumn{1}{|c|}{\vdots} & \multicolumn{1}{|c|}{\vdots} & \multicolumn{1}{|c|}{\vdots} & \multicolumn{1}{|c|}{\vdots} \\
    \bottomrule
    \end{tabular}
    \end{threeparttable}
\end{table*}

\begin{figure}
    \centering
    \begin{subfigure}{\linewidth}
    \includegraphics[width=0.98\linewidth]{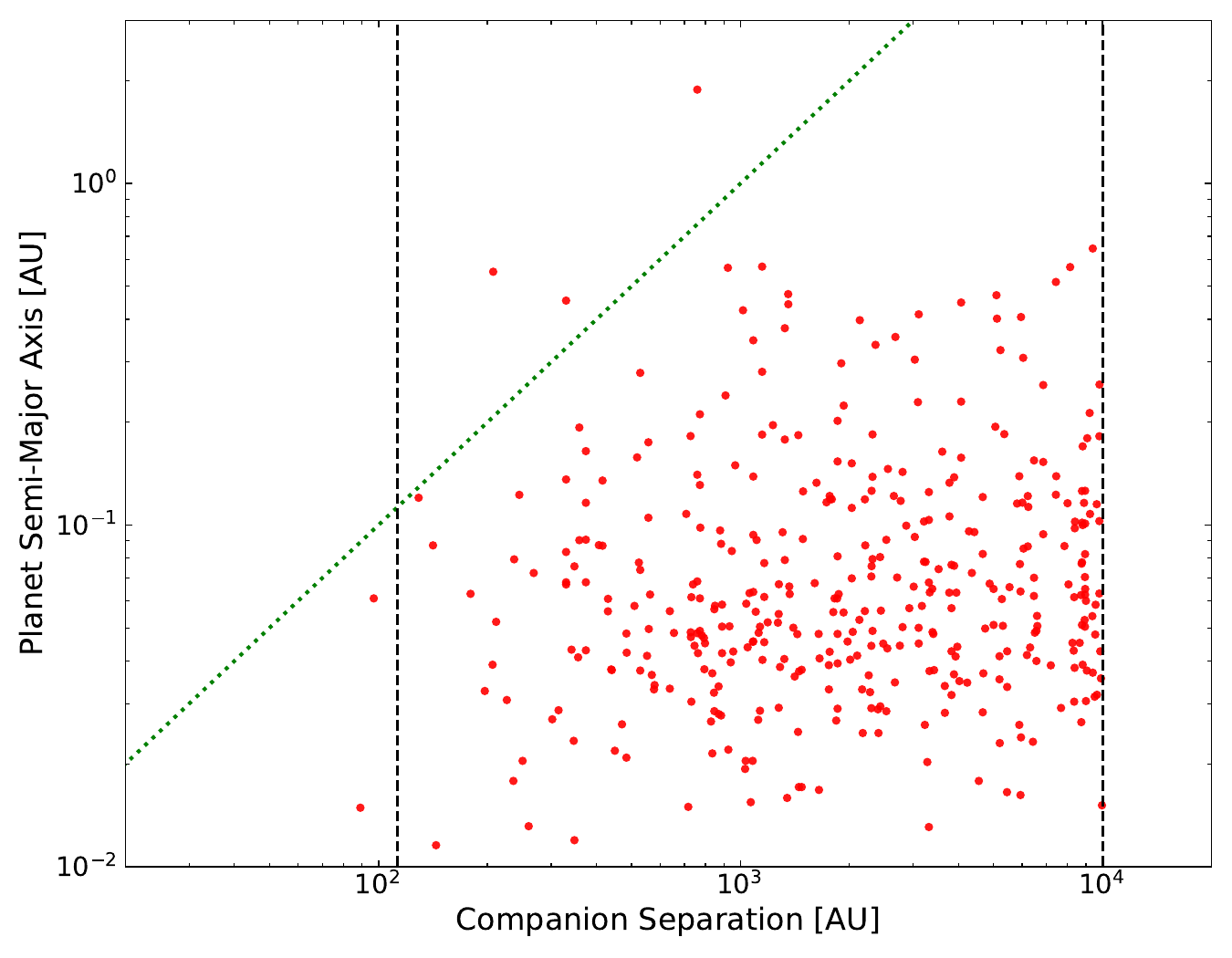}
    \end{subfigure}
    \begin{subfigure}{\linewidth}
    \includegraphics[width=0.98\linewidth]{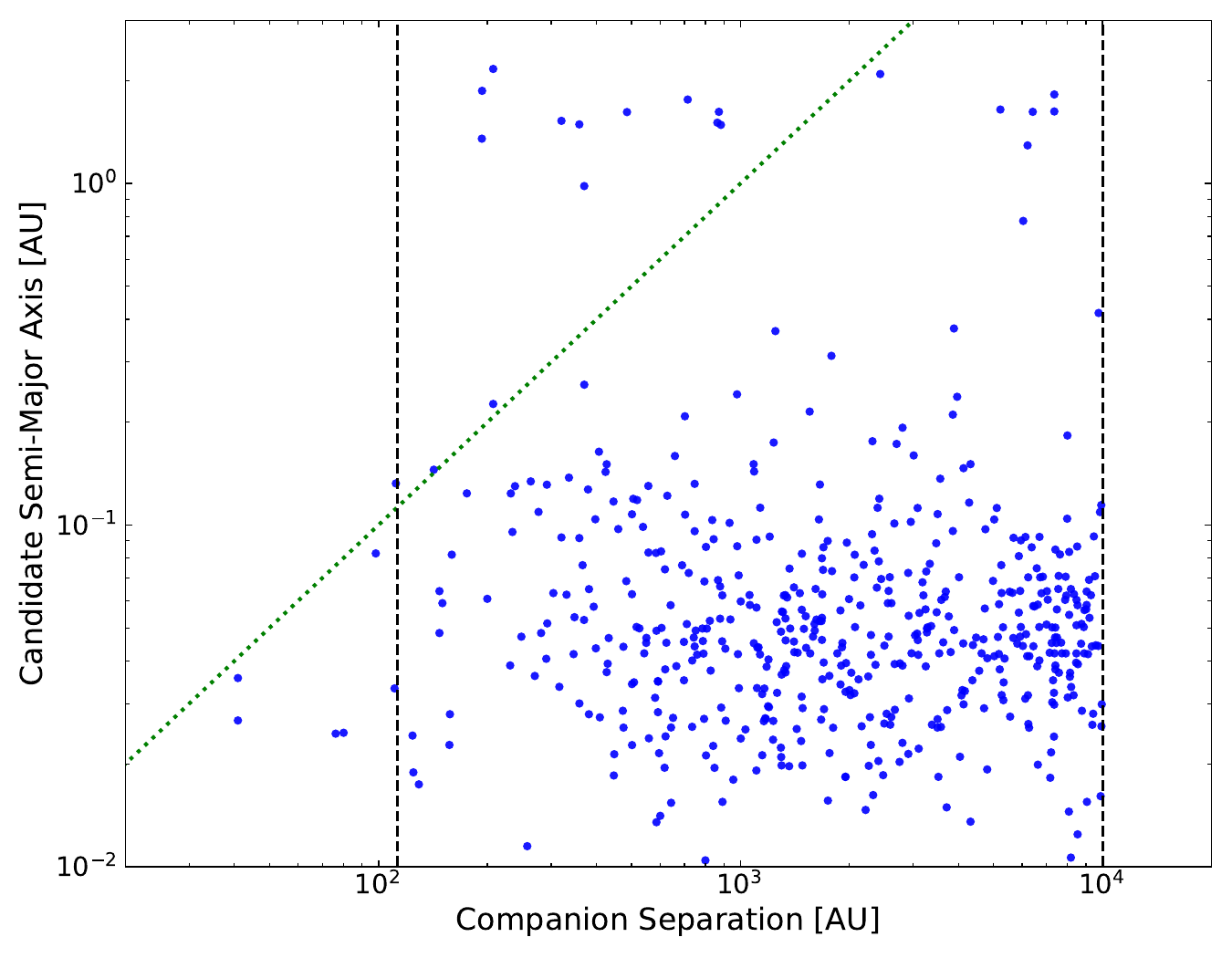}
    \end{subfigure}
    \caption{Planetary semi-major axis against companion star projected separation for confirmed (top) and candidate (bottom) exoplanets. The green dotted line corresponds to a 1000:1 stellar companion:semi-major axis relation. The black dashed lines encompass the range of stellar companion separations investigated in this work, outside of which we expect low completeness. Additional system properties measured against companion separation are available in Section \ref{sec:appendix}.}
    \label{fig:sep vs a}
\end{figure}

\begin{figure*}
    \centering
    \includegraphics[width=0.49\linewidth]{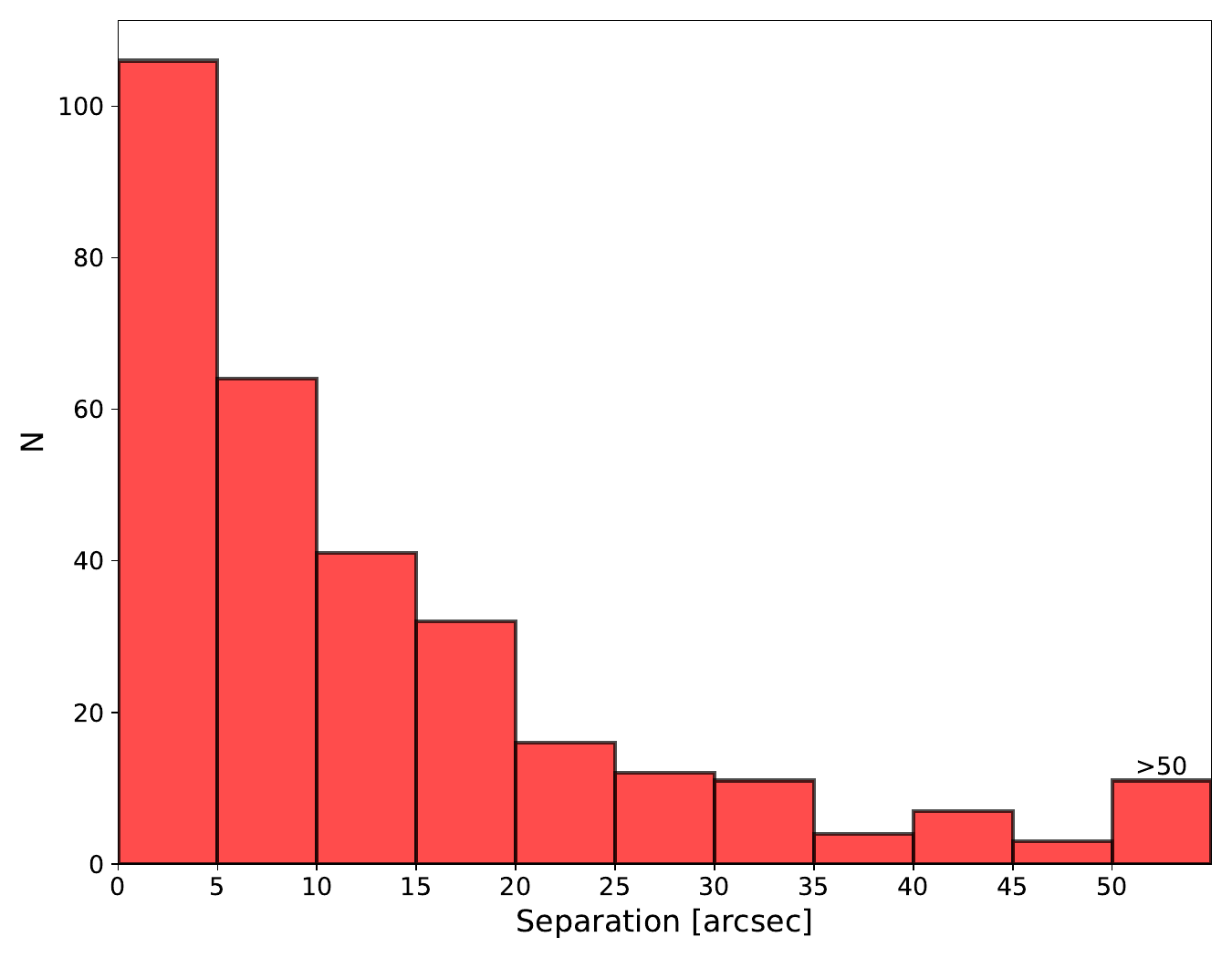}
    \includegraphics[width=0.49\linewidth]{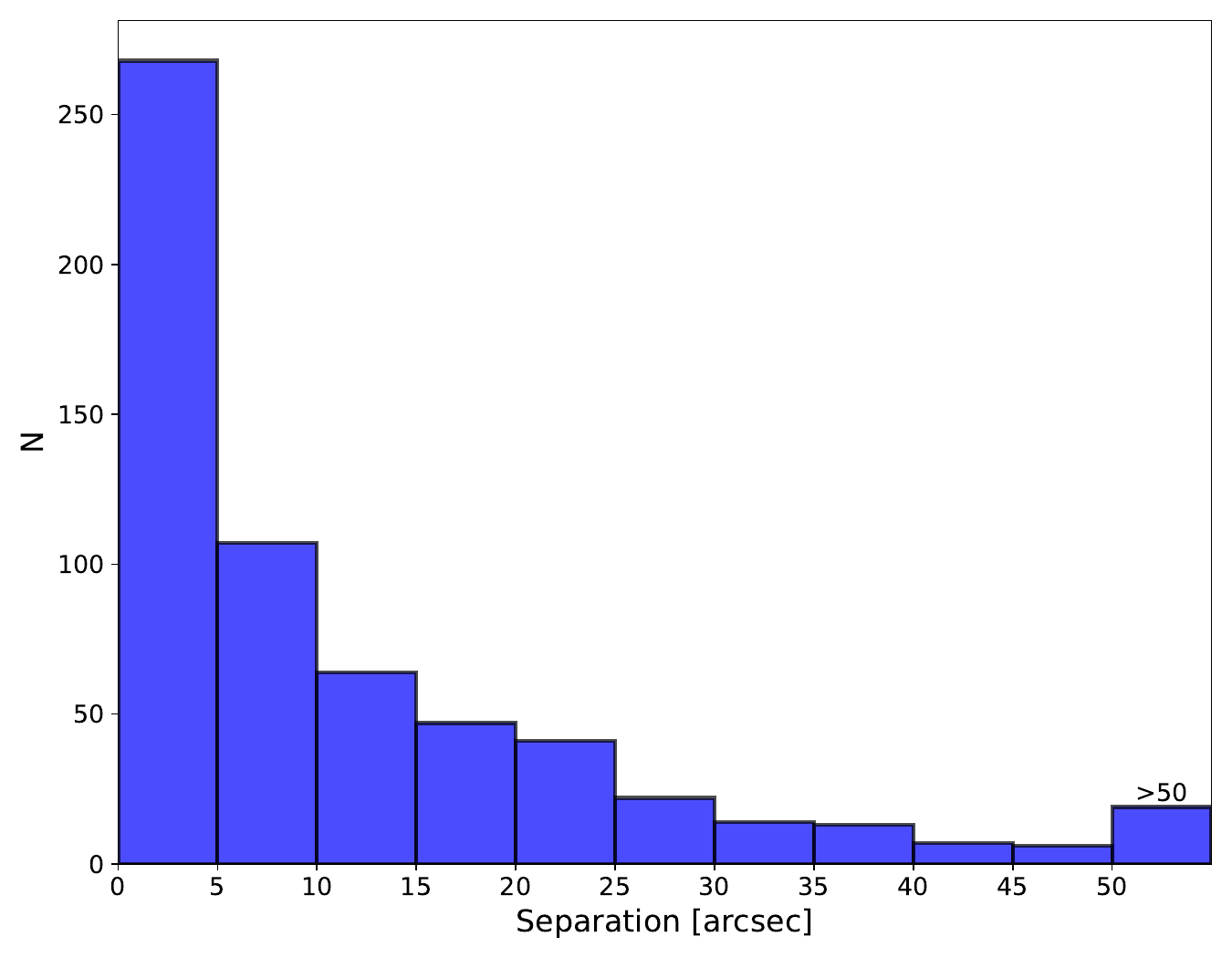}
    \caption{The distribution of angular separation for companions of confirmed exoplanet hosts (left) and TOI host stars (right).}
    \label{fig:companion sep}
\end{figure*}

The median angular separation of companions is 8.4 and 6.5 arcsec, for the confirmed exoplanet and TOI planet candidate samples, respectively. The distribution of companion separations for both samples are shown in Fig. \ref{fig:companion sep}. Using the distance to the host star calculated by the {\it Gaia} GSP-Phot module (which was provided for 92\% and 85\% of the stars with companions hosting confirmed and candidate exoplanets, respectively) we estimated the projected companion separations. In the cases where there was no distance measurement from GSP-Phot, we used the $\pi^{-1}$[arcsec] approximation. As most of our sample has distances < 500 pc, the approximated stellar distances should not require large corrections, although the bias introduced may result in an overestimation of the distances for the most distant systems \citep{bailer-jones2018, luri2018}. These projected separations are then plotted against the semi-major axis of the exoplanets (or candidates) in the multi-stellar systems in Fig. \ref{fig:sep vs a}. 60\% of the sample of confirmed exoplanets with stellar companions had an existing semi-major axis value in the Exoplanet Archive. For the rest of the sample, we use the planet's orbital period and host star mass to approximate the semi-major axis using Kepler's Third Law, assuming circular orbits. ExoFOP does not provide semi-major axis values for exoplanet candidates, so we use the third law approximation for the entire sample of exoplanet candidates. The upper completeness limit is simply the 10,000 AU radius of our initial cone search. The lower limit is an approximation of the 0.4 arcsec spatial resolution for close binaries in {\it Gaia} \citep{lindegren2018, lindegren2021}, projected at the median distance of the samples. As the companion separations that can be probed are distance dependent, the regions of lower companion separation are less complete than the outer regions. We see in Fig. \ref{fig:sep vs a} that the majority of the samples lie at stellar companion separation:semi-major axis ratios larger than 1,000:1. This is largely a product of selecting transiting confirmed and candidate exoplanets which are significantly biased towards shorter orbital periods.

The allowed parallax difference for companions with very precise {\it Gaia} astrometry (Equation \ref{eq:parallax diff allowed}) results in 2 additional multi-stellar systems hosting confirmed exoplanets; LTT 3780 and 55 Cnc, and 3 systems hosting TOI PCs; TIC 177308364, TIC 199712572 and TIC 233617847. As expected, the 2 systems with confirmed exoplanets have distances < 25 pc, however the 3 TOI systems are found much further away, with distances > 100 pc. The companions to these systems were found to have high-quality astrometry, most notably parallax significance values > 250, much higher than the median of $\sim38$. Therefore, despite the large distances to the host stars, we should be able to identify inconsistent parallax values to 3$\sigma$.

\begin{figure*}
    \centering
    \includegraphics[width=0.49\linewidth]{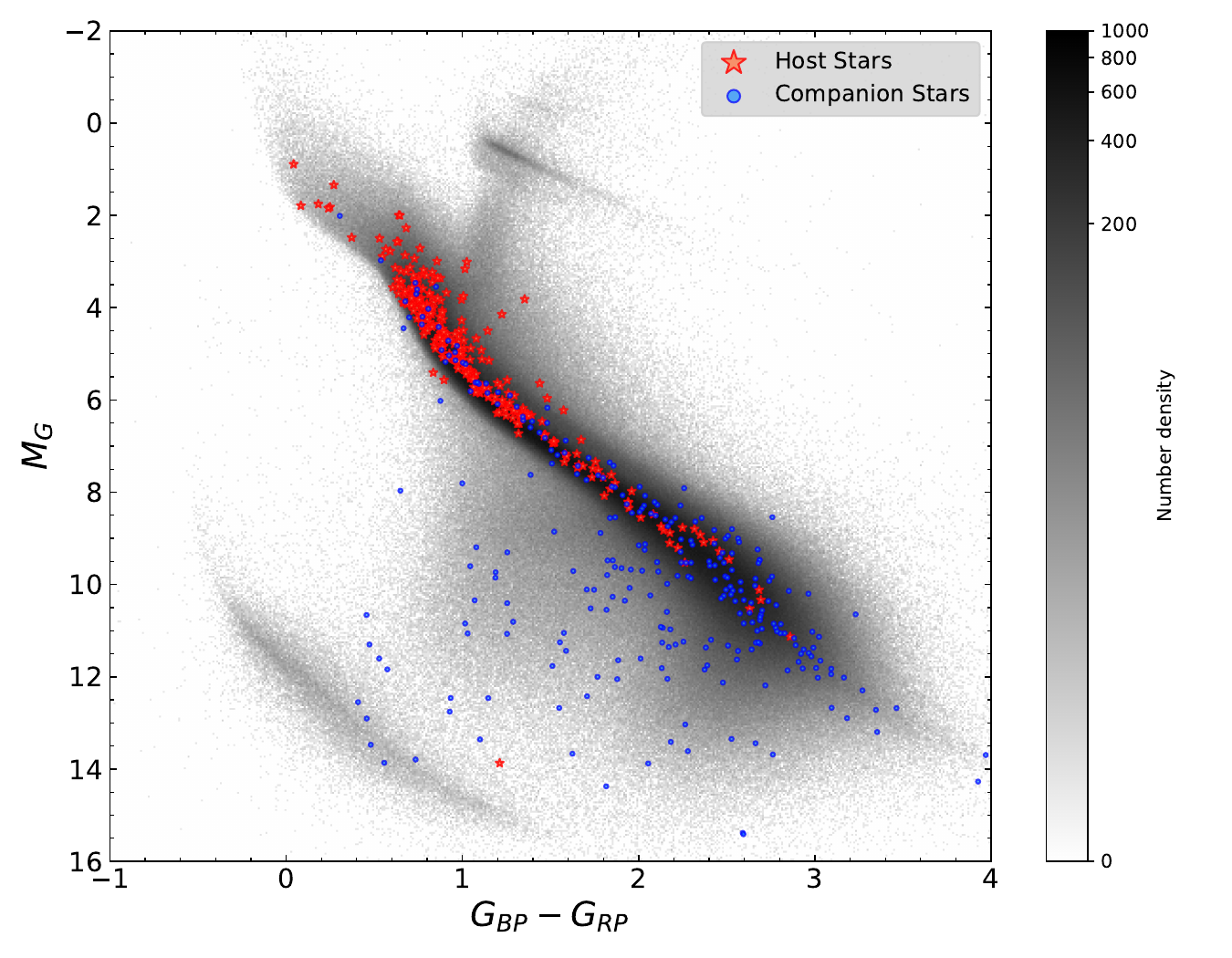}
    \includegraphics[width=0.49\linewidth]{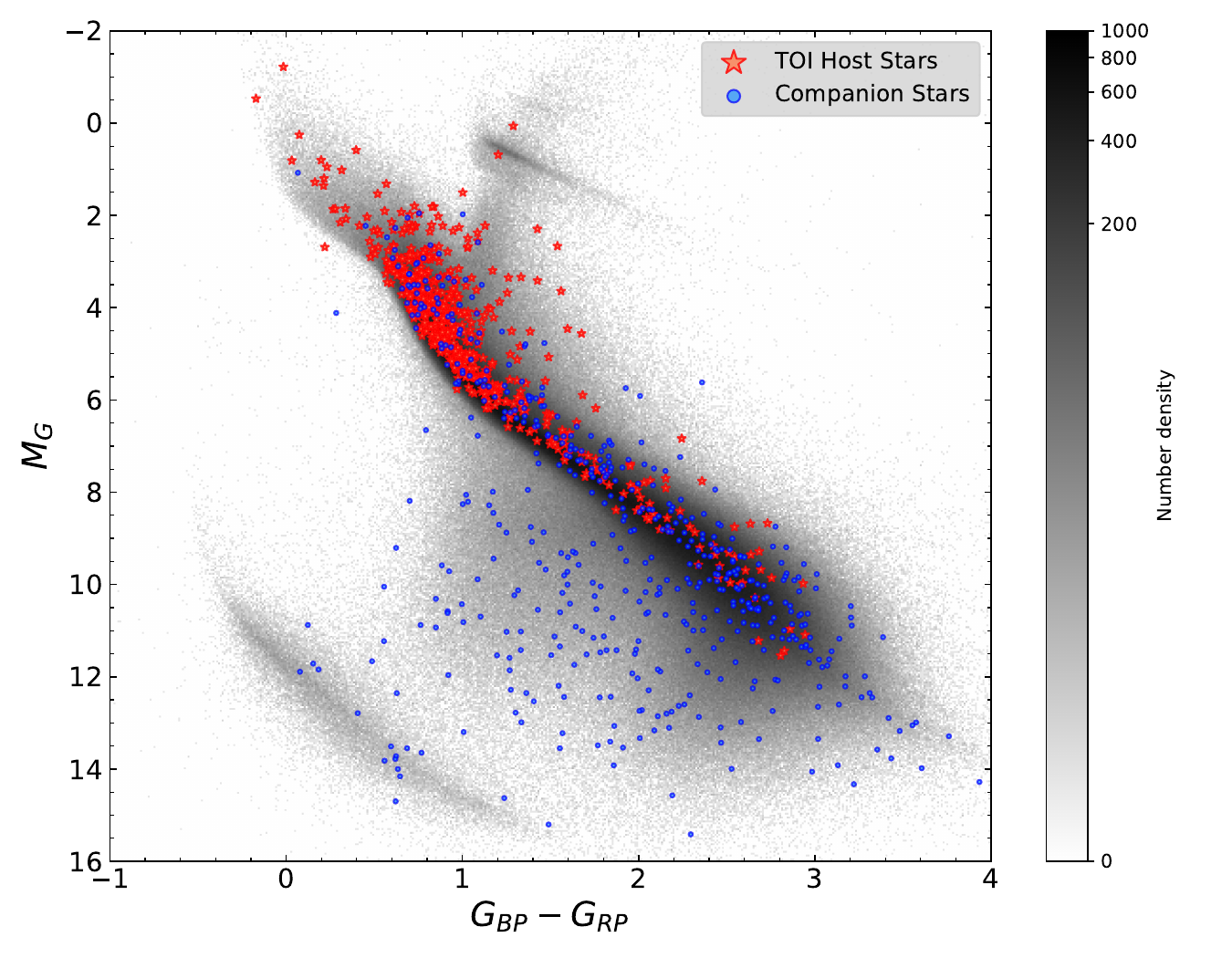}
    \caption{The colour-magnitude diagram of the CPM companion stars to host stars of confirmed exoplanets (left) and TOI PCs (right), corrected for G band extinction. The background consists of the primary and secondary stars in the \citet{el-badry2021} binary catalog. The companion sources below the main sequence are discussed in Section \ref{sec:below ms}.}
    \label{fig:colour-mag}
\end{figure*}

The positions of the host stars and their stellar companions on the {\it Gaia} DR3 colour-magnitude diagram are shown in Fig. \ref{fig:colour-mag}. The absolute G-band magnitudes of the host and companion stars have been corrected for extinction. This was achieved by searching for the star in the \texttt{StarHorse2} catalogue \citep{anders2022}. If a star had no G-band extinction listed in \texttt{StarHorse2}, we searched the \texttt{StarHorse} catalogue \citep{anders2019} instead. For stars without a G-band extinction value in either catalogue, the value of the corresponding host/companion star is adopted. If a value for extinction does not exist in either catalogue for both the host star and companion(s), we adopt the G-band extinction estimates from the \textit{Gaia} DR3 GSP-Phot module \citep{creevey2023}. For systems that do not have any estimate of G-band extinction, the value is calculated using $A_{G} / A_{V} = 0.789$ \citep{wang2019}, where the V-band extinction estimates are found in the literature, listed in the TIC \citep{stassun2019} or in the \texttt{ViZieR} data base \citep{ochsenbein2000}. The host star is used as the priority in this search.

There are 285 and 532 systems with a single stellar companion hosting confirmed exoplanets and TOI PCs respectively. 11 exoplanet systems and 38 TOI PC systems have more than one stellar companion. In total, we find 296 confirmed exoplanet hosts and 579 TOI PC hosts with detected companion sources. We find the overall stellar multiplicity rate of the confirmed exoplanet host stars in our search to be $16.6 \pm 0.9\%$. For TOIs, we find a stellar multiplicity rate of $19.8\pm0.7\%$.

\begin{figure*}
    \centering
    \includegraphics[width=0.49\linewidth]{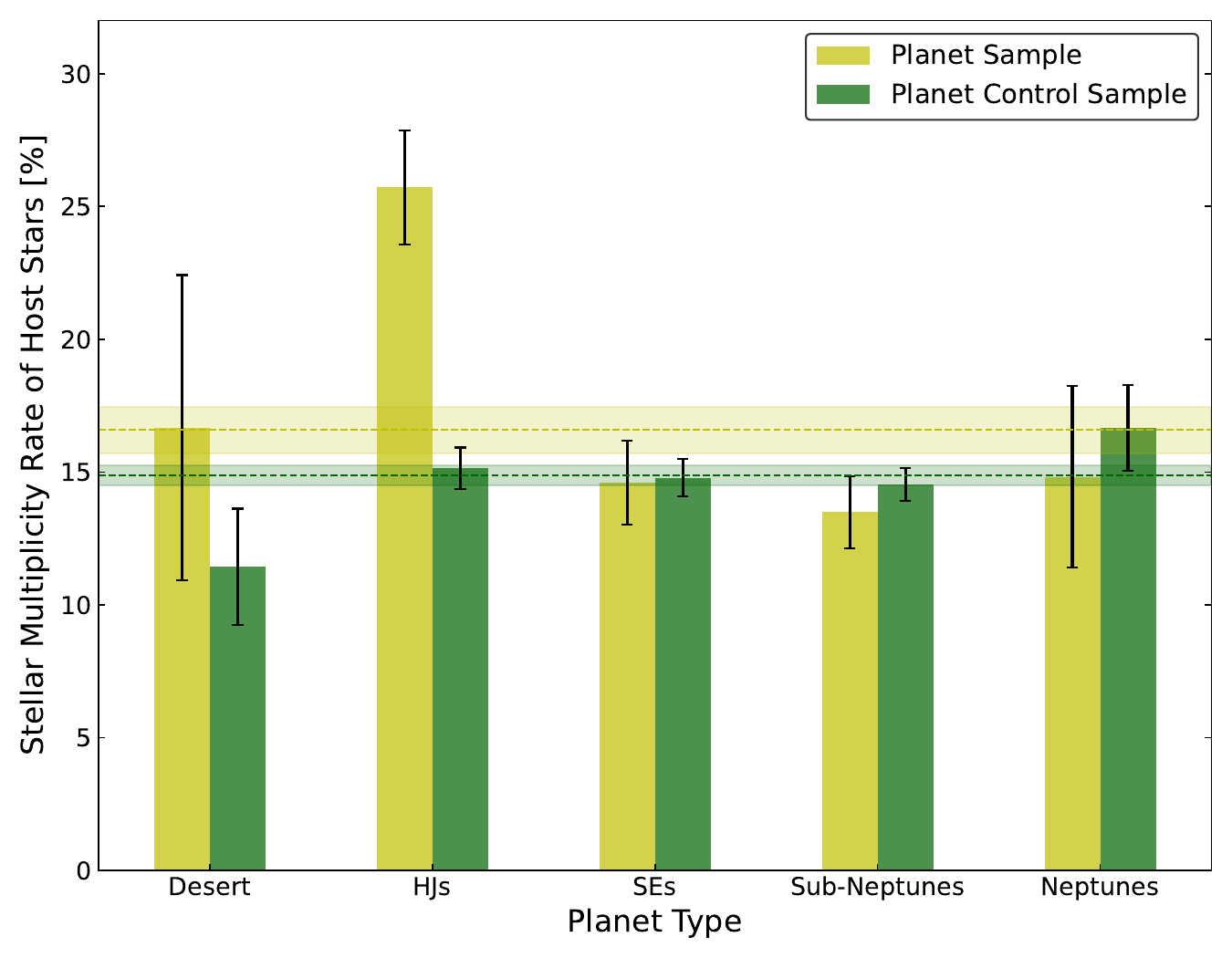}
    \includegraphics[width=0.49\linewidth]{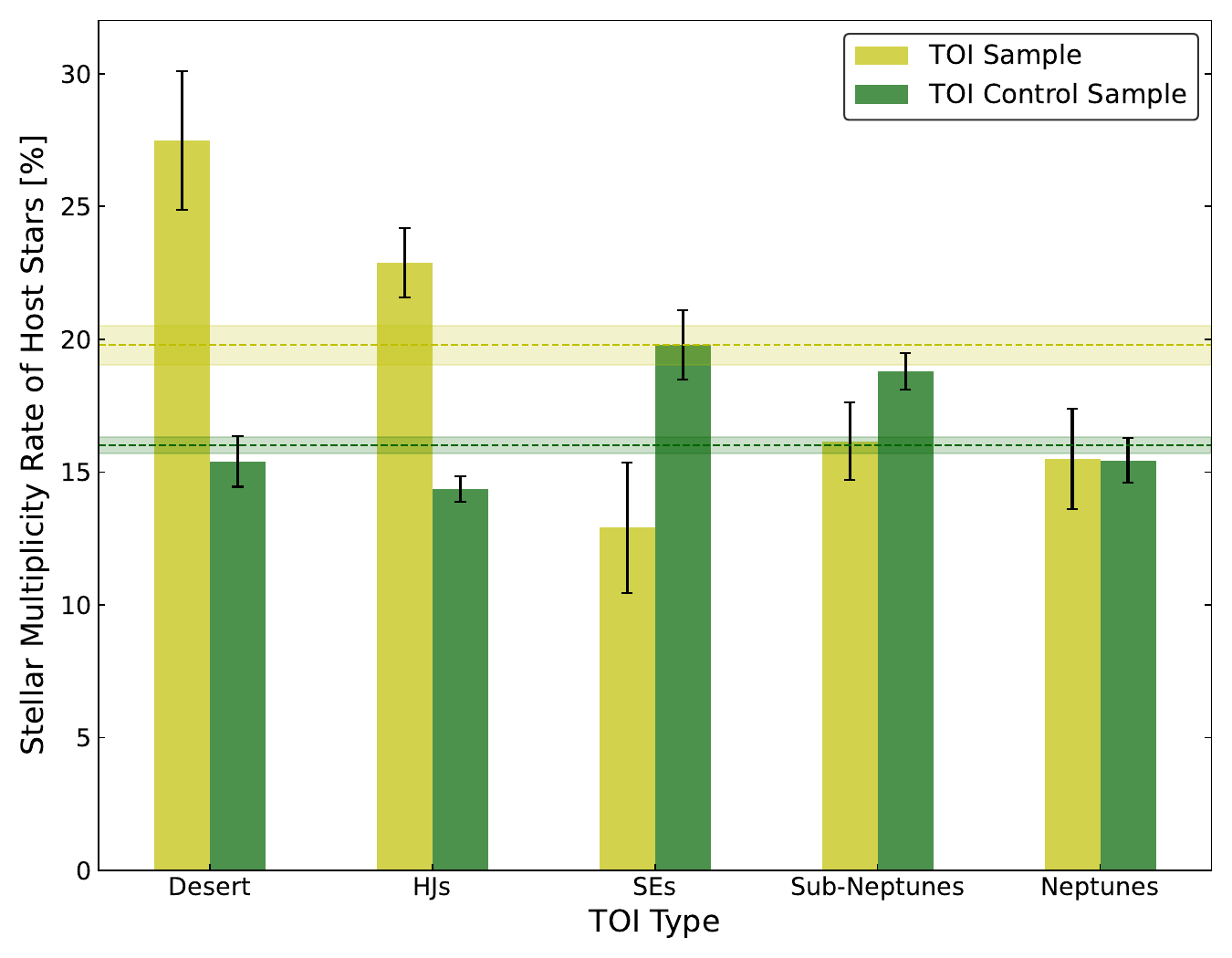}
    \caption{The stellar multiplicity rates of stars hosting confirmed exoplanets (left) and TOIs (right). The sample of exoplanets/TOIs is divided into the planet types discussed in Section \ref{sec:planet types}: Neptunian Desert, Hot Jupiters, Super-Earths, Sub-Neptunes and Neptunes. The control sample result for each exoplanet type is included for comparison. The error bars shown are the 1$\sigma$ sampling errors assuming the samples can be approximated as binomial distributions. The dashed horizontal lines are the stellar multiplicity rates of the full samples, with the shaded areas representing the sampling errors. These multiplicity rates (and the uncertainties) are provided in Table \ref{tab:rates}.}
    \label{fig:rates}
\end{figure*}

We can then separate our search samples into the planetary types discussed in Section \ref{sec:planet types}. The resulting stellar multiplicity rates of stars hosting (candidate) exoplanets of each type are listed in Table \ref{tab:rates} and plotted in Fig. \ref{fig:rates}, where the corresponding control samples are also included for comparison. We calculate the uncertainties on the stellar multiplicity rates by assuming Poisson statistics on our counts of multi-stellar systems, which then gives the $1\sigma$ confidence interval:
\begin{equation*}
    \pm \frac{1}{\sqrt{N}}\sqrt{\frac{n_{m}}{N}\left(1 - \frac{n_{m}}{N}\right)} \,
\end{equation*}
where $N$ is the total number of host stars in the (sub)sample and $n_{m}$ is the number of host stars with companions in that (sub)sample.

\begin{table}
    \scriptsize
    \centering
    \caption{Stellar multiplicity rates for different types of confirmed and candidate exoplanets. The number of multi-stellar systems, $n_{m}$, and the total number of systems, $N$, are included for each subsample. The values in brackets are the results found using the corresponding control samples, described in Section \ref{sec:control}.}
    \begin{tabular}{|l|ccc|ccc|}
    \toprule
    \multicolumn{1}{|c|}{Planet} & \multicolumn{3}{c|}{Confirmed Exoplanets} & \multicolumn{3}{c|}{TOIs}\\
    \multicolumn{1}{|c|}{Type} & Rate (\%) & $n_{m}$ & $N$ & Rate (\%) & $n_{m}$ & $N$ \\
    \midrule
    \textbf{Full Sample} & $\mathbf{16.6\pm0.9}$ & \textbf{295} & \textbf{1779} & $\mathbf{19.8\pm0.7}$ & \textbf{579} & \textbf{2927} \\
     & \textbf{($\mathbf{14.9\pm0.4}$)} & \textbf{(1322)} & \textbf{(8885)} & \textbf{($\mathbf{16.0\pm0.3}$)} & \textbf{(2339)} & \textbf{(14610)} \\
    Neptune Desert & $16.7\pm5.8$ & 7 & 42 & $27.5\pm2.6$ & 80 & 291 \\
     & ($11.4\pm2.2$) & (24) & (210) & ($15.4\pm0.9$) & (224) & (1455) \\
    Hot Jupiters & $25.8 \pm 2.1$ & 108 & 420 & $22.9\pm1.3$ & 237 & 1036 \\
     & ($15.1 \pm 0.8$) & (318) & (2100) & ($14.4\pm0.5$) & (743) & (5175) \\
    Super-Earths & $14.7\pm1.6$ & 73 & 500 & $12.9\pm2.5$ & 24 & 186 \\
     & ($14.8\pm0.7$) & (369) & (2495) & ($19.8\pm1.3$) & (184) & (930) \\
    Sub-Neptunes & $13.5\pm1.4$ & 85 & 630 & $16.2\pm1.5$ & 102 & 631 \\
     & ($14.5\pm0.6$) & (457) & (3145) & ($18.8\pm0.7$) & (593) & (3155) \\
    Neptunes & $14.8\pm3.4$ & 16 & 108 & $15.5\pm1.9$ & 57 & 368 \\
     & ($16.7\pm1.6$)  & (90) & (540) & ($15.4\pm0.8$) & (284) & (1840) \\
    \bottomrule
    \end{tabular}
    \label{tab:rates}
\end{table}

For Hot Jupiters, we find stars that host exoplanets have a stellar multiplicity rate $8.6\pm2.8\%$ higher than the corresponding control sample. This significant difference ($\sim3.1\sigma$) follows the results from \citet{ngo2016} and \citet{michel+mugrauer2023}, indicating that Hot Jupiters are more likely to have a stellar companion than other types of exoplanet.

We find a stellar multiplicity rate $7.1\pm6.2\%$ higher for stars hosting exoplanets in the Neptunian Desert compared to the control sample. Consequently, we cannot confirm any significant difference in stellar multiplicity between stars hosting planets in the Neptunian desert and the general stellar population. The small sample sizes of the Neptunian desert subsample resulted in the large uncertainty. Table \ref{tab:rates} shows how small this subsample is relative to the other exoplanet types, with a sample size of only 42 (out of 1779) exoplanet hosts. 

The multiplicity rates for Super-Earths, Sub-Neptunes and Neptune-like exoplanets were consistent with those of the corresponding control sample stars, implying stellar multiplicity has a negligible effect on the formation and evolution of these exoplanets.

Separating the TOI sample into exoplanet types, we find results similar to the sample of confirmed exoplanets when comparing the search and control samples. The exception is stars hosting planet candidates in the Neptunian desert. Although both the search sample and control sample multiplicity rates increased, the TOIs in the Desert have a stellar multiplicity rate $12.1\pm2.8\%$ higher than the control sample - a 4.3$\sigma$ significance. The difference between the stellar multiplicity rate of Hot Jupiter planet candidates and the control sample is also more significant than it was for confirmed exoplanets, with the search sample giving a multiplicity rate $8.5\pm1.4\%$ higher than the control sample ($>6\sigma$). This improved significance is a result of, relative to the sample of confirmed exoplanets, the much larger sample sizes of exoplanet candidates in these regions of parameter space. In the Neptunian Desert there were 291 TOIs in our search sample, a sample size nearly 7 times larger than the confirmed exoplanet sample.

As was the case for confirmed exoplanets, the multiplicity rates for Sub-Neptunes and Neptune-like exoplanet candidates were consistent with the rate of their control samples. However, the Super-Earth TOI sample showed a multiplicity rate $6.9\pm2.8\%$ lower than the control sample. Although this result is not statistically significant ($\sim2.5\sigma$), this does hint at the expected scarcity of lower-mass exoplanets in multi-star systems \citep{michel+mugrauer2023}.

While it is important to compare the stellar multiplicity rates of different exoplanet types with that of the full sample, comparisons between the different planet types is the more revealing result. The stellar multiplicity rate for the full sample is inclusive of all exoplanet types, thus in such a comparison there will be overlap between the samples. As different search samples in the literature (and within this work) are composed of different numbers of each type of exoplanet, the full sample stellar multiplicity rates are affected by compositional variation, and thus are not the most useful point of reference for subsample comparison. The effect can be seen using our full sample results in Table \ref{tab:rates}. Hot Jupiter are the largest subsample of planet candidates, contributing $\sim35\%$ of the full sample. Consequently, comparing the stellar multiplicity rate of Hot Jupiter planet candidates with that of the full sample of planet candidates is not as useful as the same comparison for the confirmed exoplanets, as Hot Jupiters only constitute $\sim 24\%$ of the full sample.

\subsection{Companion Sources Below the MS}\label{sec:below ms}

As shown in Fig. \ref{fig:colour-mag}, a significant portion of our CPM companions appear to exist below the main sequence in the colour-magnitude diagram. Although astrophysical objects, such as unresolved White Dwarf - M dwarf binaries \citep{el-badry2021}, may exist with these photometric properties, the majority of these sources are objects with poor quality astrometric and/or photometric data in {\it Gaia} DR3 \citep{arenou2018} and hence introduce uncertainty about their true nature.
We can assess the impact of removing a portion of these spurious companion sources on the resulting stellar multiplicity rates by applying a basic astrometric quality cut. \citet{el-badry2021} identified a cut requiring \texttt{astrometric\_sigma\_5d\_max}$ < 1$ that removed a significant portion of their initial binary star companion candidates, while only removing a small number of genuine binaries. For {\it Gaia} objects with 5-parameter astrometric solutions, \texttt{astrometric\_sigma\_5d\_max} represents the longest axis in the 5D error ellipsoid in mas. Thus, this metric is a good identifier of the general quality of astrometric data of an object, without bias towards a single measurement. Although \citet{el-badry2021} ultimately did not use this cut (opting instead to define a collection of more intricate cuts to retain the maximum number of binaries), this cut is sufficient for our purpose of a quick assessment of the effects on the stellar multiplicity rate.

We find that this cut removes $\sim11\%$ of the companions to the search sample of confirmed exoplanet hosts and $\sim9\%$ of the companions to the search sample of planet candidate hosts. Furthermore, $\sim15\%$ and $\sim13\%$ of the companions are removed for the control samples of confirmed and candidate exoplanet hosts, respectively. The overall stellar multiplicity rates were $14.7\pm0.8\%$, $17.7\pm0.7\%$, $13.2\pm0.4\%$ and $14.6\pm0.3\%$ for the search samples of confirmed and candidate exoplanet hosts, as well as the corresponding control samples, respectively. Although the stellar multiplicity rates for the individual exoplanet types change as a result of this cut, Table \ref{tab:rates cut} shows there are no significant changes to the stellar multiplicity rate difference between the search and control samples. The largest decrease was $-1.4\%$ for Neptunian Desert candidate planets. The largest increase was $+1.5\%$ for confirmed Neptune-like exoplanets. These changes are insignificant given the uncertainties involved.

Naturally, we could go further and attempt more in depth astrometric (or photometric) cuts aiming to ensure only the retention of real, well-measured {\it Gaia} stars \citep{mikkola2023}. Furthermore, we could focus on cleaning specific parameter spaces within the companion sample, such as the removal of additional chance alignments at large separations \citep{el-badry2021}. However, we did not continue with these reduced samples for the rest of this work. This is for two reasons. Firstly, even though the chosen cut does not appear to affect the relative stellar multiplicity rate of the exoplanet subsamples, our choice of cut could be introducing bias. Secondly, this allows for easier comparison with existing studies on exoplanet stellar multiplicity, which typically did not apply similar cuts.

\begin{table}
    \scriptsize
    \centering
    \caption{Stellar multiplicity rates for different types of confirmed and candidate exoplanets using the \texttt{astrometric\_sigma\_5d\_max}$ < 1$ sample cut. The values in brackets are the results found using the corresponding control samples, with the same sample cut applied.}
    \begin{tabular}{|l|c|c|}
    \toprule
    \multicolumn{1}{|c|}{Planet Type} & \multicolumn{1}{c|}{Confirmed Exoplanets Rate (\%)} & \multicolumn{1}{c|}{TOIs Rate (\%)}\\
    \midrule
    \textbf{Full Sample} & $\mathbf{14.7\pm0.8}$ & $\mathbf{17.7\pm0.7}$ \\
     & \textbf{($\mathbf{13.2\pm0.4}$)} & \textbf{($\mathbf{14.6\pm0.3}$)} \\
    Neptune Desert & $14.3\pm5.4$ & $24.7\pm2.5$ \\
     & ($10.0\pm2.1$) & ($14.0\pm0.9$) \\
    Hot Jupiters & $23.3\pm 2.1$ & $20.4\pm1.3$ \\
     & ($13.2 \pm 0.7$) & ($12.9\pm0.5$) \\
    Super-Earths & $12.2\pm1.5$ & $12.4\pm2.4$ \\
     & ($13.1\pm0.7$) & ($19.5\pm1.3$) \\
    Sub-Neptunes & $11.9\pm1.3$ & $14.9\pm1.4$ \\
     & ($13.2\pm0.6$) & ($17.2\pm0.7$) \\
    Neptunes & $14.8\pm3.4$ & $13.3\pm1.8$ \\
     & ($15.2\pm1.5$) & ($13.8\pm0.8$) \\
    \bottomrule
    \end{tabular}
    \label{tab:rates cut}
\end{table}

\section{Discussion}\label{sec:discussion}

Using astrometric and photometric data from {\it Gaia} DR3, we investigated the stellar multiplicity of a sample of 1779 stars hosting confirmed exoplanets in the NASA Exoplanet Archive. We performed the same companion search on 2927 stars that host exoplanet candidates discovered by the {\it TESS} mission. The parallaxes and proper motions of each exoplanet (candidate) hosting star were used to search for common proper motion companions, utilising the low characteristic uncertainties on {\it Gaia} DR3 astrometric data. Within our sample, we find 295 exoplanet and 579 candidate hosts with one or more companion stars. This translates to a stellar multiplicity rate of 16.6$\pm 0.9\%$ and 19.8$\pm0.7\%$ for confirmed and candidate exoplanet hosts, respectively. 

We will now compare these rates with previous works on stellar multiplicity of exoplanet systems which make use of {\it Gaia} astrometry. \citet{gonzalez-payo2024} performed a search using DR3 for visual companions with $s < 1 $ pc in a smaller volume of $d < 100$ pc and found a rate of 21.6$\pm2.9\%$ after discarding open cluster pairs and ultra-cool dwarf companions. \citet{fontanive2021} searched a $d < 200$ pc volume in DR2 with a 20,000 AU companion separation limit and found a rate of 23.2$\pm1.6\%$. \citet{michel+mugrauer2023} also used DR3 and found a rate of 16.0$\pm0.8\%$ in a volume of $d < 625$ pc before adding stellar companions from their own literature search. Using DR2, \citet{michel2021} found a rate of 16$\pm2\%$ for $d < 500$ pc. At $d < 500$ pc \citet{mugrauer2023} found a rate of 14.9$\pm1.1\%$ for (Community)TOIs, which became 19.9$\pm1.5\%$ for only TOIs (these TOIs consisted of all EXOFOP dispositions, not just planet candidates). 

Although these results are relatively consistent with each other, it should be noted that these stellar multiplicity rates are always an underestimation of the true value as a consequence of companion sample incompleteness, which will change with the volume limit. As a basic test, we ran smaller volume limits on our sample during the companion search, and we found no major impact on the relative stellar multiplicity rates of different exoplanet types. For the sample of confirmed exoplanets, the only noticeable change from steadily decreasing the volume limit in steps of 50 pc was that the stellar multiplicity rate of Hot Jupiters increased to $\sim30\%$ for smaller limits ($<250$ pc). For the sample of exoplanet candidates, the subsample of Hot Jupiters also reach a stellar multiplicity rate of $\sim30\%$ at 250 pc, and the rate for Neptunes increases as well ($\sim21\%$ at 250 pc). In all cases, the decreasing sample sizes increased the errors on the stellar multiplicity rates, such that any changes may be statistically insignificant.

The studies with smaller volume limits will be more complete at lower stellar masses (both the host star and companion), however the effect this would have on any perceived multiplicity rate is complex. Despite previous works agreeing that low mass stars have a lower multiplicity rate than brighter stars \citep{duchene2013}, M-dwarfs also account for more than 70\% of the stars in the Milky Way \citep{henry2006, winters2014}. Additionally, most stellar companions to M-dwarfs are found at separations $<50$ AU \citep{dhital2010}, a region of parameter space not investigated by these wide companion surveys.

The uncertainties on our stellar multiplicity rates were determined using the Poisson uncertainties assuming a binomial distribution on our samples (each host star is either a single or multiple stellar system).

It is likely that some of these detected stellar companions are in fact unresolved binaries. In general, companion stars are fainter and redder than the exoplanet (candidate) host star, leading to lower quality astrometric and photometric data on average. In such cases, the blending of two or more close objects becomes more likely. This is important when the companion is at a large separation to the host star, as wide triple systems are more common than wide binaries \citep{basri2006, tokovinin2006, cifuentes2021}.

Dividing the overall exoplanet (candidate) population into subsamples corresponding to mutually exclusive exoplanet types defined by orbital period and planet radius, we calculated the stellar multiplicity rates of stars hosting different types of exoplanets (candidates). To compare the multiplicity of these subsamples to that of similar stars in the field, we built control samples and performed the same stellar companion searches. These control samples consisted of non-planet-hosting stars in {\it Gaia} DR3 that shared similar astrophysical or photometric properties to the corresponding exoplanet (candidate) hosting sample. It should be noted that despite the control samples representing non-planet-hosting stars, it is impossible to ensure the removal of stars that are hosting undiscovered planets. To date, few systems have been observed in ways that probe large portions of exoplanet radius and orbital separations. Consequently, even if we were to only select stars that had been monitored as part of some previous survey and returned a null result, the likelihood of a non-detected planet would still be significant. 

We investigated any similarities between the samples of stars hosting Hot Jupiters and exoplanet (candidates) in the hot Neptunian Desert. Host stars of both confirmed exoplanets and {\it TESS} exoplanet candidates show a larger stellar multiplicity rate when the exoplanet (candidate) was a Hot Jupiter. This result was the same for stars hosting Neptunian Desert exoplanet candidates. The sample of confirmed exoplanets in the Neptunian Desert was too small to confirm any difference in the multiplicity rate relative to the corresponding control sample.

The higher-than-field stellar multiplicity rate of stars hosting both Hot Jupiter exoplanets (and candidates) and Neptunian Desert candidates is indicative of another preferential environmental condition for the formation of gas giant exoplanets with short orbital periods. However, this link is largely dependent on the true nature of the exoplanet candidates in the Neptunian Desert. Assuming these candidates are not false positives, the follow-up observations should be largely unaffected by the presence of a wide stellar companion. Minimum separations on existing spectrographs such as HARPS or CARMENES are well-known \citep{bonfils2013, quirrenbach2014, cortes-contreras2017} and are sufficient to avoid contamination from most of the stellar companions found in this work. For example, the 5 arcsec (seeing dependent) limit at CARMENES results in only companions with projected separations $< 1,000$ AU contaminating spectra of the host star at $d = 200 $ pc.

As the sample of confirmed exoplanets grows alongside an increasing range of exoplanet physical properties, it is important to identify and understand any features that may be shared by different types of exoplanets. The formation and evolution of stars, and that of any planets they host, is tightly linked to stellar multiplicity, which is a factor in more than half of Solar-type stars. Thus, the continued investigation into stellar multiplicity in the context of planetary formation and evolution is an important step in understanding the observed trends in exoplanet demographics. The confirmation and characterisation of the the many remaining exoplanet candidates in the Neptunian Desert is a priority in this regard. The release of {\it Gaia} DR4 will identify many more exoplanet candidates, as well as any stellar companions with smaller separations (thanks to epoch astrometry data).

\section{Conclusion}\label{sec:conclusion}

In this work, we built samples of stars hosting confirmed exoplanets and exoplanet candidates identified by the {\it TESS} mission at distances less than 650 pc. Using {\it Gaia} DR3 astrometry, we searched for nearby objects to stars in these samples and detected 946 sources that were found to have both similar parallaxes and common proper motions. For our total sample of confirmed exoplanet hosts, we find a stellar multiplicity rate of $16.6\pm0.9\%$. For the sample of stars hosting exoplanet candidates, we find an overall rate of $19.8\pm0.7\%$. We measure the projected separations between these stars and investigate the relation between the exoplanet semi-major axis and stellar companion separation. Plotting the positions of the host and companion stars on the {\it Gaia} DR3 colour-magnitude diagram, we find that $\sim20\%$ of the companions are found below the main-sequence. This region is associated with poor quality {\it Gaia} DR3 astrometric and/or photometric data, indicating that they may be artefacts or unresolved White-Dwarf and late-type main-sequence binaries. Although including these companions is in line with established works, we calculate the stellar multiplicity rate with and without these companions. We confirm that the increased stellar multiplicity of Hot-Jupiter and Neptunian Desert planet candidates is robust with or without these companions.

By separating the samples of confirmed and candidate exoplanets into planet types using orbital period and planet radius, we find stellar multiplicity rates of $16.7\pm5.8\%$, $25.7\pm2.1\%$, $14.6\pm1.6\%$, $13.5\pm1.4\%$ and $14.8\pm3.4\%$ for exoplanets in the hot Neptunian Desert, Hot Jupiters, Sub-Neptunes, Super-Earths and Neptune-like planets, respectively. The equivalent rates for candidate exoplanets are $27.5\pm2.6\%$, $22.9\pm1.3\%$, $12.9\pm2.5\%$, $16.2\pm1.5\%$ and $15.5\pm1.9\%$.

Using a combination of photometry and GSP-Phot stellar parameters, we built control samples consisting of randomly selected stars at similar distances and of similar spectral types to the those in both our search samples. By performing a similar companion search on these control samples (which are $5\times$ larger than the search samples), we can identify the differences in stellar multiplicity rate of stars hosting (candidate) exoplanets and the general population, without introducing bias from the strong multiplicity dependence on spectral type. We find that stars with Neptune-like planets, Sub-Neptunes and Super-Earths appear to have similar stellar multiplicity rates as the general population. However, stars hosting Hot Jupiters or Neptunian Desert (candidate) exoplanets have significantly higher stellar multiplicity rates than similar field stars. This result for Hot Jupiters is in agreement with previous multiplicity studies, but it was unknown whether this trend continued into the Neptunian mass regime at similar orbital periods. Although this result has $<3\sigma$ significance in the confirmed exoplanet sample (a result of the relatively small sample size), the result in the {\it TESS} exoplanet candidate sample may be evidence of another environmental feature shared between Hot Jupiters and planets in the Neptunian Desert.

Looking to the future, analysis of the many {\it TESS} planetary candidates used in this work should be prioritised. Through the confirmation (or rejection) of exoplanet candidates in the Neptunian Desert, the relative error on the stellar multiplicity rate will be reduced, and one can assess whether the higher rate observed in the candidate sample is shared by the confirmed exoplanets. However, even as the Neptunian Desert sample size increases, the stellar multiplicity rate will likely not increase to the level seen in the planet candidates due to the assumed existence of false positives.

Adaptive imaging and radial velocity surveys complement wide companion searches such as this work well. They probe much lower projected separations than those available in {\it Gaia}, improving the sample completeness at separations < 100 AU. Combining catalogs of planet-hosting multi-star systems with well-understood biases will be the first step in the creation of a homogeneous sample of companion objects to host stars, allowing the fundamental effects of stellar multiplicity on exoplanet formation and evolution to be investigated.

\section*{Acknowledgements}

We thank the anonymous referee for their constructive comments. This research was funded by the UKRI (Grants ST/X001121/1, EP/X027562/1). This research has made use of the NASA Exoplanet Archive, which is operated by the California Institute of Technology, under contract with the National Aeronautics and Space Administration under the Exoplanet Exploration Program. This research has made use of the Exoplanet Follow-up Observation Program (ExoFOP; DOI: 10.26134/ExoFOP5) website, which is operated by the California Institute of Technology, under contract with the National Aeronautics and Space Administration under the Exoplanet Exploration Program. This paper includes data collected with the {\it TESS} mission, obtained from the MAST data archive at the Space Telescope Science Institute (STScI). Funding for the {\it TESS} mission is provided by the NASA Explorer Program. STScI is operated by the Association of Universities for Research in Astronomy, Inc., under NASA contract NAS 5–26555.

This work has made use of data from the European Space Agency (ESA) mission {\it Gaia} (\url{https://www.cosmos.esa.int/gaia}), processed by the {\it Gaia} Data Processing and Analysis Consortium (DPAC, \url{https://www.cosmos.esa.int/web/gaia/dpac/consortium}). Funding for the DPAC has been provided by national institutions, in particular the institutions participating in the {\it Gaia} Multilateral Agreement. 

This research has made use of the SIMBAD database, VizieR catalogue access tool (DOI : 10.26093/cds/vizier) and cross-match service, all operated at the CDS, Strasbourg Astronomical Observatory, France. This research made use of {\tt Astropy}\footnote{\url{http://www.astropy.org}}, a community-developed core Python package for Astronomy \citep{astropy2013, astropy2018} and {\tt Astroquery}\footnote{\url{https://astroquery.readthedocs.io/}} \citep{ginsburg2019}.

%%%%%%%%%%%%%%%%%%%%%%%%%%%%%%%%%%%%%%%%%%%%%%%%%%
\section*{Data Availability}

Full versions of tables in this paper are available online.



\bibliographystyle{mnras}
\bibliography{multiplicity} % if your bibtex file is called example.bib

%\begin{thebibliography}{99}
%\end{thebibliography}

%%%%%%%%%%%%%%%%%%%%%%%%%%%%%%%%%%%%%%%%%%%%%%%%%%

%%%%%%%%%%%%%%%%% APPENDICES %%%%%%%%%%%%%%%%%%%%%

\appendix

\section{Additional Figures} \label{sec:appendix}

\begin{figure*}
    \centering
    \begin{subfigure}{\textwidth}
    \includegraphics[width=0.98\linewidth]{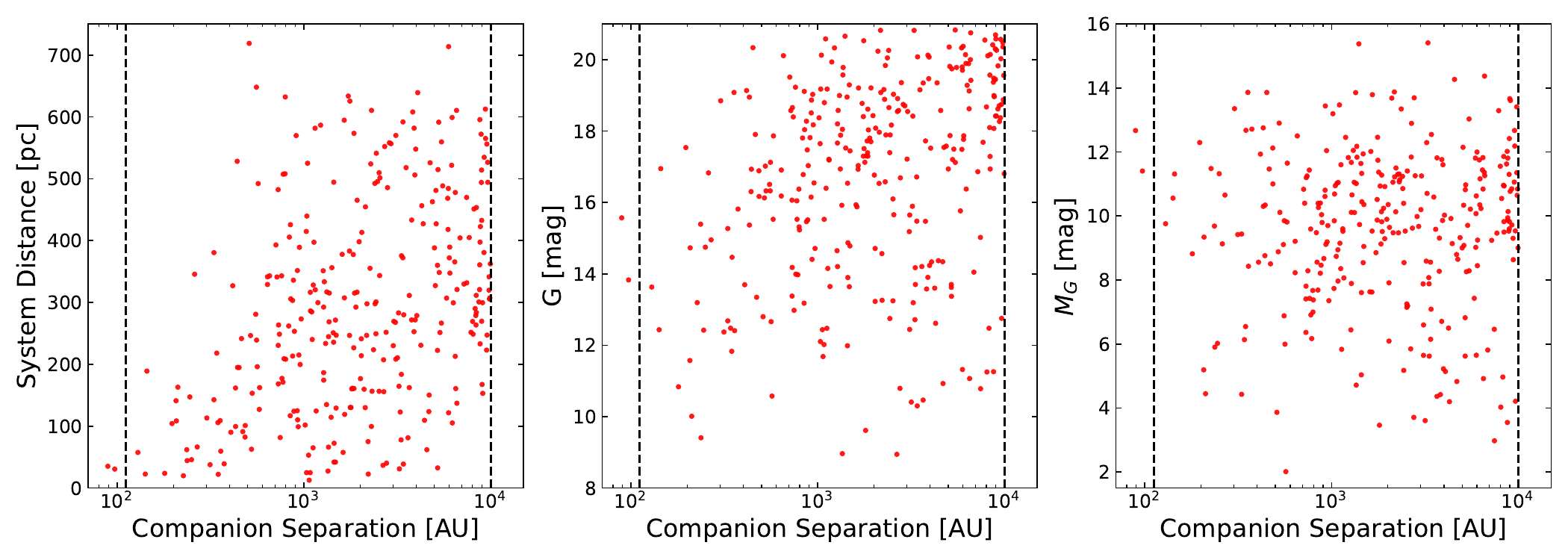}
    \end{subfigure}
    \begin{subfigure}{\textwidth}
    \includegraphics[width=0.98\linewidth]{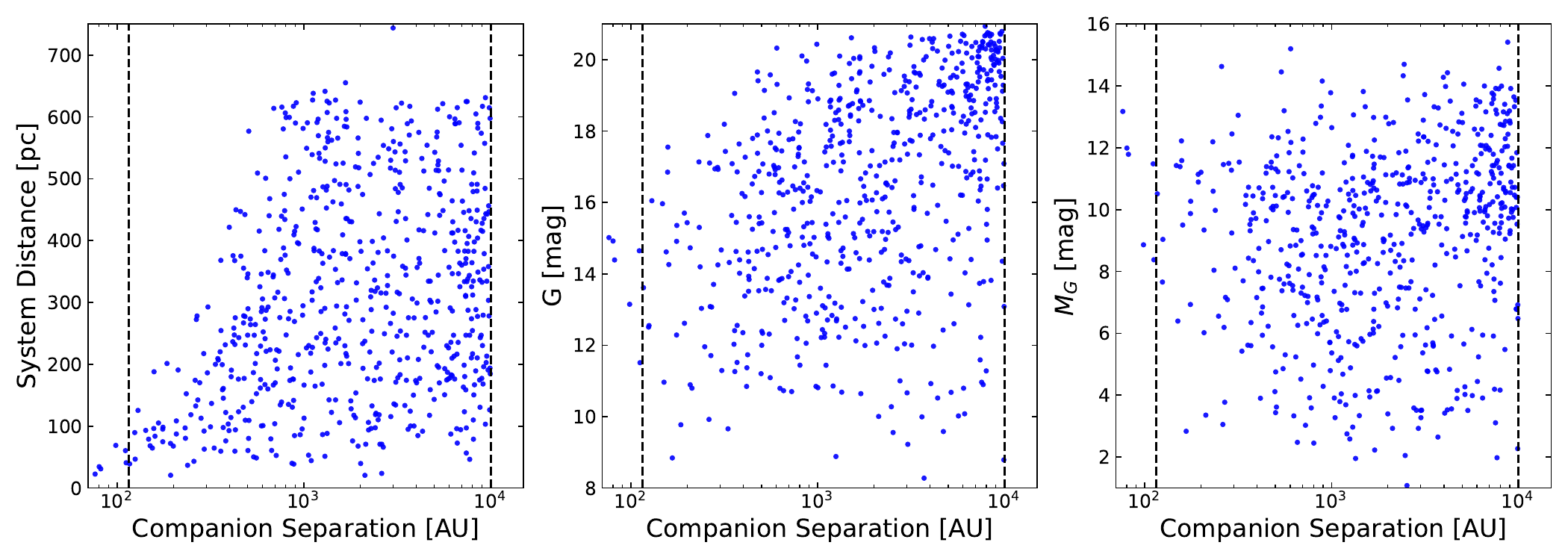}
    \end{subfigure}
    \caption{Companion separation against system distance (left), apparent G magnitude (centre) and extinction-corrected absolute G magnitude (right) of the companion stars to confirmed exoplanets (top) and planet candidates (bottom). \textit{\textbf{Note:} The single system with distance > 650 pc in the lower left plot is the result of a host star with high parallax uncertainty, but is still within our significance limits.}}
    \label{fig:comp sep plots}
\end{figure*}

%%%%%%%%%%%%%%%%%%%%%%%%%%%%%%%%%%%%%%%%%%%%%%%%%%

% Don't change these lines
\bsp	% typesetting comment
\label{lastpage}
\end{document}